\documentclass[aps,prl,reprint,nofootinbib,amsmath,amssymb,notitlepage,]{revtex4-2}

\usepackage{graphicx}
\usepackage{dcolumn}
\usepackage{bm}
\usepackage{url}
\usepackage[hyperindex,breaklinks]{hyperref}
\usepackage{placeins} 
\usepackage{xcolor}
\usepackage{xparse}

\hypersetup{linkcolor=blue, citecolor=red, colorlinks=true}

\NewDocumentCommand{\tens}{t_}
 {%
  \IfBooleanTF{#1}
   {\tensop}
   {\otimes}%
 }
\NewDocumentCommand{\Log}{o}{%
  \IfNoValueTF{#1}{}{{}^{#1}\!}\log}%

\newcommand{\arcsinh}{\mathrm{arcsinh}}
\newcommand{\sech}{\mathrm{sech}}
\newcommand{\csch}{\mathrm{csch}}

\begin{document}

\title{Statistics of stochastic entropy for recorded transitions between ENSO states}

\author{S. M. \surname{Duarte~Queir\'{o}s}}
\thanks{sdqueiro@cbpf.br}

\affiliation{
Centro Brasileiro de Pesquisas F\'isicas\\
Rua Dr Xavier Sigaud, 150, 22290-180, Urca, Rio de Janeiro -- RJ, Brazil
}

\date{\today}

\begin{abstract}
We analyse the transitions between established phases of the El Ni\~no Southern Oscillation (ENSO) by surveying the daily data of the Southern Oscillation Index from an entropic viewpoint using the framework of stochastic Statistical Physics. We evaluate the variation of entropy produced due to each recorded path of that index during each transition as well as taking only into consideration the beginning and the end of the change between phases and verified both integral fluctuation relations. The statistical results show that these entropy variations have not been extreme entropic events; only the transition between the strong $1999-2000$ La Ni\~na to the moderate $2002-2003$ El Ni\~no is at the edge of being so. With that, the present work opens a long and winding avenue of research over the application of stochastic Statistical Physics to Climate Dynamics.
\end{abstract}

                              
\maketitle

From chit-chat in a lift to ground breaking Science paving the way to the Nobel Prize~\cite{parisi1982}, climate is a hot topic. Besides the climatic issues linked to humankind (in)actions that have been in the spotlight, climate dynamics has its own regular phenomena that create important distress in our lives~\cite{carleton2016}. Among them, the El Ni\~ no (EN) and La Ni\~ na (LN) episodes are certainly at the first-tier~\cite{cashin2016}. Each event corresponds to the opposing phases of the ENSO 
first conveyed by G.~Walker on his study over the `Southern Oscillation'~\cite{walker1924}: large-scale changes of the sea level pressure across Southeast Asia and the tropical Pacific. Nonetheless, it took 45 years~\cite{bjerknes1969} to assert a connection between ENSO and changes in the ocean, namely the perception by South American fishermen of the warming up of coastal waters that occurred every so often around Christmas.

The EN corresponds to a warming/above-average temperature of the ocean surface in the central and eastern tropical Pacific~\cite{l'heureux2014}; the precipitation over Southeast Asia decreases while rainfall increases over the tropical Pacific. At the same time, easterly low-level surface winds along the equator weaken or even turn westerly instead. 
Conversely, during the LN, the ocean surface experiences a cooling/below-average temperature and the weather effect over Southeast Asia and central tropical Pacific is opposite to that of EN; in addition, the typical equatorial easterly winds become stronger. In-between, there is the so-called Neutral phase in which the tropical Pacific surface temperature is close to its average or, more often than not, ocean conditions match a given 
`child'~\footnote{El Ni\~no and La Ni\~na literally translate from Castilian into `The Boy' and `The Girl', respectively.} 
state, but the not the atmosphere or the other way round.

Despite the fact that EN/LN are coupled (ocean and atmosphere) phenomenon~\cite{l'heureux2014,behringer1998,ji1998,trenberth2007,conti2013}, the identification climate is getting in or out of one of them is widely related to the Southern Oscillation Index (SOI) that measures the intensity of the Walker Circulation: the driven tropical atmospheric circulation in the longitudinal direction~\cite{bjerknes1969}. It defines a measure of the difference in the surface air pressure (anomaly) between Tahiti and Darwin expressed in standard deviation units~\cite{soi-formula}. Although commonly presented in a monthly timeframe, SOI daily values are accessible since $1991$. The rule-of-thumb linking SOI and ENSO phases goes as follows~\cite{bureau2012}: prolonged (average) positive SOI values above $+8$ indicate a LN event, whereas continuing negative values below $-8$ indicate an EN phase. In Tab.~\ref{episodes}, we indicate the 11  fully established transitions according to the World Meteorological Organization~\cite{wmo} since daily records are available.
\vspace{-0.625cm}
\begin{table}[h]  
\caption{Transitions between catalogued El Ni\~no $\leftrightarrow$ La Ni\~na ENSO events comprising the neutral phases as well.}  
\centering  
\begin{tabular}{c c c c c c c} 
\hline\hline   
 nr: direction && Start Date  && End Date 
\\   
\hline   
1: EN$\rightarrow$LN && 31st August 1992 && 1st September 1995 \\
2: LN$\rightarrow$EN && 31st March 1996 && 1st March 1997 \\
3: EN$\rightarrow$LN && 31st August 1998 && 1st June 1999 \\
4: LN$\rightarrow$EN && 31st May 2000 && 1st March 2002\\
5: EN$\rightarrow$LN && 28th February 2003 && 1st June 2007\\
6: LN$\rightarrow$EN && 28th February 2008 && 1st June 2009\\
7: EN$\rightarrow$LN && 31st May 2010 && 1st June 2010\\
8: LN$\rightarrow$EN && 31st May 2011 &&  1st March 2014 \\
9: EN$\rightarrow$LN && 31st May 2016 && 1st September 2017 \\
10: LN$\rightarrow$EN && 31st May 2018 && 1st September 2018\\
11: EN$\rightarrow$LN && 31st May 2019 && 1st June 2020\\
\hline   
\end{tabular} 
\label{episodes}
\end{table} 
\vspace{-0.375cm}

The dynamics of ENSO/SOI has been mimicked manifold: from assuming a non-linear systems approach~\cite{chang1996,choi2013,bruun2017} to a stochastic perspective of the problem~\cite{ausloos2007,jin2007a,jin2007b,an2020,kim2020}. 
Praising all the others, we rank the latter as particularly suited to set forward an analysis within stochastic statistical physics (SSP)~\cite{sekimoto2010,seifert2012}. Therein, the description of physical laws hinges on the probability distribution, $p( \{\mathcal{O} _t\})$, of a trajectory assumed by the system -- related to the observable $\mathcal{O}$ -- in going from a state into another and recurrently presented in the form of (integral) fluctuation relations~\cite{galavotti2008}. An instance of that is the Jarzynski relation~\cite{jarzynski1997}, which can be understood as the probabilistic version of the Clausius inequality~\cite{jarzynski2011}. Moreover, this fluctuation oriented description lifted the veil over the different components of the entropy identifying its informational contribution associated with the intrinsic modification of the probability and those related to energetic quantities, viz, heat~\cite{esposito2010,sasa2001,seifert2005}.
Although by heeding the thermodynamic limit, we would expect such fluctuations were only meaningful for small systems~\cite{klages2013}, the truth is that it is possible to set a probabilistic description, namely for entropy production, of systems as big as wind-tunnel experiments under fully developed turbulence~\cite{fuchs2020} or rogue wave statistics~\cite{lind2018}. Thence, even if we are treating the quintessential macroscopic system (the atmosphere), it is plausible that in carrying out a SSP analysis of the SOI we can still obtain relevant insights, namely the probabilistic features of the trajectories taken by that index when ENSO goes from state into the other. At this point, natural questions arise: {\it owing that EN and LN are the contrasting phases of ENSO, does the blatant difference in the respective typical weather (i.e. heat/thermal-entropy) and patterns have found quantitative correspondence in the entropy variations derived from the SOI dynamics? 
Do they correspond to extreme events as one is readily prone to assert since they imply extreme weather and heat phenomena? What further can we learn by employing SSP to climate?} The answer to these questions constitutes our goal.

\smallskip
\smallskip

To learn over the dynamics of SOI, $\{ s _t \}$, we use its daily series recorded between the 157th day of 1991 and 212th day of 2023~\cite{soi-data}. These data are jointly listed and computed from the pressure measurements used to get the monthly SOI, which is more convenient to appraise trends due to the natural reduction of the fluctuations. Still, in the trajectory approach, we aim at being as close as possible to the actual SOI path; thus, we focus on the analysis of the smallest available sampling rate the monthly form would wipe out valuable knowledge. For all that, within SPP~\cite{klages2013}, it is often assumed a  standard differential formulation containing a time-dependent contribution, $\Pi (t)$, representing a driving process,
%
$
d\mathcal{O}/dt = f(\mathcal{O}) + \Pi (t) + \sqrt{2\, g(\mathcal{O})}\, \eta _t$, 
%
where $\left\langle \eta_t \, \eta _{t^\prime} \right\rangle = \delta (t \, - t^\prime )$ in a Stratonovich representation. Milestone findings in SSP~\cite{cohen2003} considered, $\Pi (t) = \mathrm{const} \,t$, yielding, $\mathcal{O}_t = \mathrm{const} \,t + \xi _t$, which concentrates the stochasticity on $ \xi _t$. For the SOI, a similar description is possible and relates to the usual geophysical procedure of describing a trend ruled by some non-linear  $\Pi (t)$ though. Willing to keep close to the methods usually employed in ENSO analysis, a candidate to infer the SOI protocol, $\Phi (t)$, is the use of the Multi-Taper Method~\cite{thomson1982,ghil2002} used in several other areas as well~\cite{karnik2022}. Following that~\cite{ghil2002,mann1996,ausloos2008}, we applied the method for $K=5$ tapers and the subsequent spectral analysis indicated as statistically significant the set of frequencies yielding,
\begin{math}
\{T\}= \{24, 28, 36, 74, 102, 365, 2168 \}
\end{math} 
days, a set that matches several geophysical and astronomical phenomena related to ENSO~\cite{SM-1}. The inversion to the time domain of the so-called background dynamics -- ie, protocol $\Phi $ -- was carried out conforming to the geophysical signal processing technique~\cite{ghil2002,mann1996} of a nearly optimal reconstruction by means a mean-square minimisation of the numerical adjustment of the reconstructed signal with respect to $  \Phi (t) = \Phi_0(t)+ \sum _{i=1}^7  A_i(t) \cos \left[ \frac{2\, \pi}{T_i}t+\theta_i \right]$; the details and further values are presented in~\cite{SM-1}. Thence, we define,
\begin{math}
\xi _t \equiv s_t - \Phi _t\end{math}, 
as stationary following the generic stochastic equation,
\begin{math}
d\xi = D_1 ^\prime (\xi) \, dt + \sqrt{2\, D_2 ^\prime (\xi)}\, dW_t.
\end{math}
From the data we evaluate the empirical Kramers-Moyal coefficients~\cite{risken1989},
%
$
\tilde{D}_n ^\prime (\xi,t) \equiv \frac{1}{t\, n!} \left\langle (\xi(t) - \xi )^n \right\rangle _{ss},
$
%
($ss$ stands for averages assuming a stationary state). For $t \rightarrow 0$, we have $\tilde{D} ^\prime _n(\xi ,t) = D_n ^\prime (\xi)$ ($D_n ^\prime (\xi) =0$, for $n \ge 3$).
%
%
Following Ref.~\cite{anteneodo2010}, we incorporate the SOI sampling rate, $\tau = 1$ day, and in the It\^o definition we get (details in SM),
\vspace{-0.25cm}
\begin{equation}
D_1 ^\prime (\xi) = - a\, \xi + b, \qquad D_2 ^\prime (\xi) =  \alpha ^2 \, \xi ^2 + \beta \, \xi + \gamma ^2,
\label{km-xi}
\vspace{-0.15cm}
\end{equation}
where $a = 0.15 \pm 0.01$, $b = 0.08 \pm 0.01$, $\alpha ^2 = 0.010 \pm 0.002 $, $\beta = -0.14 \pm 0.03$ and $\gamma ^2 = 31.65 \pm 0.8$  [see Figs.~3~a) and~b) in the SM]. To further probe the reliability of $\xi $ dynamics, we generated $\xi $ series numerically and compared $\tilde D_3 ^\prime (\xi)$ and $\tilde D_4 ^\prime (\xi)$ with that found to the SOI $\xi $component. The agreement corroborates the stochastic model defined by Eq.~(\ref{km-xi}). Blending both $\xi _t$  and $\Phi _t$  dynamics we get,
\vspace{-0.2cm}
\begin{equation}
ds = D_1(s,t) \, dt + \sqrt{2\, D_2(s,t)}\, dW_t,
\label{langevin-soi}
\vspace{-0.2cm}
\end{equation}
(It\^o representation of the noise) with,
\vspace{-0.2cm}
\begin{equation}
D_1(s,t) = - a\, s + a\, \Phi (t) + b, \,\, D_2(s,t) =  \alpha _s ^2 \, s ^2 + \beta _s \, s + \gamma _s ^2,
\vspace{-0.2cm}
\end{equation}
where,
$\beta _s = \beta - 2 \, \alpha ^2\, \Phi(t)$ and $\gamma _s ^2 = \gamma ^2 + \alpha ^2 \, \Phi ^2 (t) - \beta \, \Phi (t)$.

As previously stated, we primarily want to evaluate the variation of entropy, $ \Delta S (\vec s =\{s_{t_i}, \ldots, s_{t_f} \}) $~\cite{spinney2013}, 
\vspace{-0.1cm}
\begin{eqnarray}
\Delta S (\vec s) &\equiv &- \ln \frac{p(s_{t_f};\{ s_{t_i}, \ldots, s_{t_{f-1}}\})}{p(s_{t_i};\{ s_{t_{i+1}} \ldots s_{t_f}\})} \label{entropy} \\
 &=&  - \ln \frac{p_{t_i}^*(s_{t_i}) \, p^F(s_{t_{i+1}}|s_{t_i},\Phi) \ldots p^F(s_{t_f}|s_{t_{f-1}}, \Phi )}{p_{t_f}^*(s_{t_f}) \, p^R(s_{t_{f-1}}|s_{t_f},\Phi ) \ldots p^R(s_{t_i} | s_{t_{i+1}},\Phi )}, 
\nonumber
\vspace{-0.2cm}
\end{eqnarray}
($F$ and $R$ stand for forward and reverse trajectory) produced in the evolution of SOI along the trajectory segment starting at day $t_i$ and ending at day $t_f$ that took ENSO from a given phase into its opposite with realised final and initial probability distributions. Using the framework of SSP, the `Southern Oscillation' {\it protocol}, $\Phi (t)$, acts on the system and its outcome is its evolution from a given (local) stationary state, $p_{t_i}^*(s)$ into a new state, $p_{t_f}^*(s)$~\cite{spinney2013}, where $p_{t}^*(s)$ is obtained by considering that at time $t$ the probability current, $j_{t} (s)$, vanishes,
\vspace{-0.2cm}
\begin{eqnarray}
 j_{t}^*(s)&=& D_1(s,t) \,  p_{t}^*(s) -  \frac{\partial }{\partial x} D_2(s,t) \,  p_{t}^*(s)   = 0   
 ,
\vspace{-0.2cm}
\end{eqnarray}
or, $ \frac{\partial j_{t}^*(s)}{\partial x}  =  - \frac{\partial p_{t}^*(s)}{\partial t} = 0$, which yields,
\begin{eqnarray}
p_{t}^*(s)   & = & \frac{1}{\mathcal{Z}^* _t} \exp \left[ \frac{a\,\beta+2\,b\,\alpha ^2}{\alpha ^2 \sqrt{4\alpha ^2 \, \gamma ^2 - \beta ^2}} 
\arctan \left[ \frac{\beta+2 \, \alpha ^2 (s-\Phi(t))}{\sqrt{4\alpha ^2 \, \gamma ^2 - \beta ^2}}   \right]
\right]
\times \nonumber \\
&& \times \left[\gamma ^2 + \beta (s- \Phi (t)) + \alpha ^2 (s - \Phi(t))^2 \right]^{- \left(1+\frac{a}{2\, \alpha ^2} \right)} .
\label{prob-stat}
 \end{eqnarray}
 
To evaluate Eq.~(\ref{entropy}) we still need to compute the conditioned probabilities $p^{F(R)}(s_{t ^\prime}|s_{t},\Phi )$. To that, we employ a path integral formulation~\cite{wio2013}, but first some points must be handled. 
First, we tackle the multiplicative noise nature of the SOI dynamics that is also found in other atmospheric quantities~\cite{sardeshmukh2005}. Several techniques were cast at providing a probabilistic solution to multiplicative noise systems, but they fundamentally boil down to transformations of variables, parameters and expansions~\cite{cugliandolo2010,wio2013,ao2014,lozano2016,barci2019}. In order to maintain the work as straightforward as possible, we opt to transform the original multiplicative noise dynamics into an additive noise one as a means to retrieve $p(s_{t ^\prime}|s_{t},\Phi )$ at the end.
That is worked at in the SM leading to the transformation of variables~\cite{risken1989}
$
y= \frac{1}{\alpha} \arcsinh \left[ \frac{\alpha}{\lambda} \left(s+\frac{\beta _s}{2 \, \alpha ^2} \right)   \right],
$
%
where $\lambda ^2 = \gamma ^2 - \frac{\beta }{4 \, \alpha ^2} $.
Second, we must remove the impact of the trajectory direction on the stochastic dynamics caused by the It\^o interpretation of noise in Eq.~(\ref{langevin-soi}). That is made by rewriting Eq.~(\ref{langevin-soi}) into its Stratonovich version~\cite{gardiner1997}. Combining both steps it finally yields,
\begin{equation}
\frac{dy}{dt} = \left(\frac{b}{\lambda} +\frac{a\, \beta}{2\, \lambda \, \alpha ^2} \right)  \sech [\alpha \, y]  -   \left(\alpha + \frac{a}{\alpha} \right) \tanh [\alpha \, y]
+ \sqrt{2} \, \eta _t .
\label{langevin-y}
\end{equation}
The derivation of $p(y,t|y_0,t_0)$ from Eq.~(\ref{langevin-y}) equals~\cite{wio2013},
\begin{equation}
p(y,t|y_0,t_0) = \int \mathcal{D}y \, \exp \left[ -\frac{1}{4} \mathcal{S}[y(t)]] \right]  \, \det \frac{\delta \eta _t}{\delta y_0},
\label{pathintegral}
\end{equation}
assuming the optimising condition, $\delta \, \mathcal{S}[y(t)]] = 0$, of the stochastic action,
%
$\mathcal{S}[y(t)]  \equiv  \int _{t_0} ^t \mathcal{L}(t^\prime) dt^\prime,$
%
where the Onsager-Machlup function reads~\cite{SM-2},
\begin{equation}
\mathcal{L} =  \left\{ \dot{y} - \left(\frac{b}{\lambda} +\frac{a\, \beta}{2\, \lambda \, \alpha ^2} \right)  \sech [\alpha \, y]  +   \left(\alpha + \frac{a}{\alpha} \right) \tanh [\alpha \, y] \right\} ^2,
\label{onsager-machlup}
\end{equation}
allowing the application of a saddle point approximation.

The optimisation is achieved by solving the Euler-Lagrange equation,
%
$
\left( \frac{\partial \mathcal{L}}{\partial y}-\frac{d}{dt}\frac{\partial \mathcal{L}}{\partial \dot {y}} \right) \Big\vert _{y=y^*} = 0.
$
%
Because such equation is not analytically solvable to the best of our efforts, we focus on the parameters in Eq.~(\ref{langevin-soi}) and understand the SOI fluctuations are dominated by the additive contribution to the noise given by $\gamma ^2$, whereas the remainder multiplicative contributions act as perturbations. Expanding 
the Euler-Lagrange equation in powers of $\alpha $ and $\beta $ so that $(\alpha \, \gamma / \beta )^2 \gg 1$ --- jointly with the inspection of further relations between the parameters --- yields,
%
$
\ddot{y}^* - c_1 \, y^* + c_0 = 0,
$
%
which is bounded by $y^*(t)=y$ and $y^*(t_0)=y_0$ with $c_0 = \frac{b}{\gamma } (a+\alpha ^2)+ \frac{a\, \beta}{2\, \gamma} \left( 1+ \frac{a}{\alpha ^2} \right)$ and 
$c_1 = a^2 + 2 \, a \, \alpha ^2- (b \, \alpha ^2 + a \, \beta) b / \gamma ^2 $.
The explicit solution thereto 
is presented in the SM; utilising that solution into the stochastic action
, $\mathcal{S}$ into Eq.~(\ref{pathintegral}), and finally reverting the change of variables $y=y(s)$, 
we obtain $p^F(s(t)\,|\,s(t_0))$ and $p^R(s(t_0)\,|\,s(t))$. The (endless) formulae of both of them are shown in the SM either.

%
\begin{table}[h]  
\caption{Realised variation of the entropy, $\Delta S (\vec s)$, average, $\left\langle \Delta  S (\vec s) \right\rangle$, and standard deviation, $ \sigma _{\Delta S (\vec s) }$ over trajectories.}  
\centering  
\begin{tabular}{c c c c c c c c} 
\hline\hline   
transition nr && $ \Delta S (\vec s)$  && $\left\langle  \Delta S (\vec s) \right\rangle$ &&& $ \sigma _{ \Delta S (\vec s) }$
\\   
\hline   
1 && $1.61 \times 10^{-1}$  && $9.6 \times 10^{-5}$ &&& $8.15 \times 10^{-2}$ \\
3 && $- 6 \times 10^{-4}$  && $2.01 \times 10^{-3}$ &&& $9.05 \times 10^{-2}$ \\
5 && $ -5 \times 10^{-4}$ && $-1.17 \times 10^{-3}$ &&& $8.17\times 10^{-2}$ \\
7 && $4 \times 10^{-4}$ && $-3.6 \times 10^{-4}$ &&& $4 \times 10^{-2}$ \\
9 && $-2.9 \times 10^{-3}$ && $-2.8 \times 10^{-4}$ &&& $8.16 \times 10^{-2}$ \\
11 && $3.1 \times 10^{-2}$ && $8.6 \times 10^{-4}$ &&& $7.99 \times 10^{-2}$ \\
2 && $-1.553 \times 10^{-1}$ && $-6.9 \times 10^{-4}$ &&& $8.65 \times 10^{-2}$ \\
4 && $3.30 \times 10^{-1}$ && $3.1 \times 10^{-4}$ &&& $8.13 \times 10^{-2}$ \\
6 && $8 \times 10^{-4}$ && $1.11 \times 10^{-3}$ &&& $8.74\times 10^{-2}$ \\
8 && $- 5.3 \times 10^{-3}$ && $2.2 \times 10^{-3}$ &&& $8.33 \times 10^{-2}$ \\
10 && $-1 \times 10^{-4} $ && $8.6 \times 10^{-4}$ &&& $7.99 \times 10^{-2}$ \\
\hline   
\end{tabular} 
\label{pathentropy}
\end{table} 
At last, we compute Eq.~(\ref{entropy}) considering the trajectories of the transitions in Table~\ref{pathentropy} as well as the average, $\left\langle S (\vec s) \right\rangle$, and the standard deviation, $ \sigma _{\Delta S (\vec s) }$ over trajectories given by Eq.~(\ref{langevin-soi}). From Table~\ref{pathentropy}, we learn that only $3$ out of the $11$ transitions have $| \Delta S (\vec s) |$ above $2\, \sigma $, namely transitions nr~1,~2,~and~4; only the last of these cases --- a transition from strong LN to moderate EN~\cite{goldengate} --- is a $4\, \sigma $ event and could statistically be considered an extreme event. Even so, 
all the transitions abide by the Integral Fluctuation Theorem, $\left\langle \exp \left[ - \Delta  S \right] \right\rangle = 1$, a probabilistic version of the 2nd Law of Thermodynamics $\Delta S \ge 0$~\cite{jarzynski2011}.

\begin{table}[h!]  
\caption{Realised variation of the entropy, $\Delta \tilde S (t_f,t_i)$, 
the average, $\left\langle \Delta \tilde S (t_f,t_i) \right\rangle$, and standard deviation, $ \sigma _{\Delta \tilde S (t_f,t_i) }$, over trajectories.}  
\centering  
\begin{tabular}{c c c c c c c c} 
\hline\hline   
transition nr && $\Delta \tilde S (t_f,t_i)$  && $\left\langle \Delta \tilde S (t_f,t_i) \right\rangle$ &&& $ \sigma _{\Delta \tilde S (t_f,t_i) }$
\\   
\hline   
1 && $1 \times 10^{-4}$  && $5.8 \times 10^{-4}$ &&& $9.64 \times 10^{-2}$ \\
3 && $1.78 \times 10^{-3}$  && $-9 \times 10^{-6}$ &&& $9.85 \times 10^{-2}$ \\
5 && $1.9 \times 10^{-4}$ && $7.3 \times 10^{-4}$ &&& $9.98 \times 10^{-2}$ \\
7 && $3.6 \times 10^{-4}$ && $-1.0 \times 10^{-3}$ &&& $5.52 \times 10^{-2}$ \\
9 && $1.75 \times 10^{-3}$ && $5.3 \times 10^{-4}$ &&& $1.01 \times 10^{-1}$ \\
11 && $5.35 \times 10^{-3}$ && $7.8 \times 10^{-4}$ &&& $9.34 \times 10^{-2}$ \\
2 && $-1.727 \times 10^{-1}$ && $-6.6 \times 10^{-4}$ &&& $1.07 \times 10^{-1}$ \\
4 && $2.622 \times 10^{-1}$ && $-5.9 \times 10^{-4}$ &&& $1.0 \times 10^{-1}$ \\
6 && $3.04 \times 10^{-3}$ && $-1.55 \times 10^{-3}$ &&& $8.82\times 10^{-2}$ \\
8 && $2.24 \times 10^{-3}$ && $-1 \times 10^{-5}$ &&& $1.07 \times 10^{-1}$ \\
10 && $1 \times 10^{-4}$ && $-1.2 \times 10^{-4}$ &&& $9.11 \times 10^{-2}$ \\
\hline   
\end{tabular} 
\label{startendentropy}
\end{table}

Still within SPP, our problem can be surveyed in a slightly different way by strictly looking at the variation of entropy,
\begin{math}
\Delta \tilde S (t_f,t_i)  \equiv \int ds_{t_i+1} \ldots ds_{t_f-1} \,  \Delta S (\vec s).
\end{math}
The respective outcomes are exhibited in Table~\ref{startendentropy}. Therefrom, we perceive all but events nr~2 and nr~4 have $|\Delta \tilde S |$ less than $\sigma / 10 $, whereas for the former  $|\Delta \tilde S |$ is less than $2\sigma$ and for the latter is less than $3\sigma$. Herein, we verify the Integral Fluctuation Theorem as well. 
Regarding these results, we cannot assert transition nr~4 is an extreme event as in specific trajectory approach.

\smallskip 

In this manuscript, we have aimed at learning whether the acute difference in the typical weather and patterns of EN and LN phases of ENSO is mirrored in a cornerstone quantity such as the variations of entropy the SOI experiences when climate evolves from one phase into the other, as it would be expected in a phenomenon-indicator relation. We have done so considering the SSP framework and asserting to the SOI a stochastic dynamics evolving under a climate protocol that drives the changes in the system. To the best of our knowledge, this work inaugurates the application of such mindset to climate variables. We have proceeded by assuming the SOI path between phases as provided by its daily records and also considering  its values at the boundaries of each transition. {\it  The results indicate the EN$\rightarrow$LN transitions have had quite regular values of entropy variation during the process, usually a less  than $\sigma /10$-event. For LN$\rightarrow$EN transitions we have found more noteworthy values, namely the transitions nr~2 and nr~4 in Table~\ref{episodes}, which is a $4\, \sigma$ event in the first  approach and a $3\, \sigma$ event in the second one though}. All said and done, the transition from the strong $1999-2000$ LN to the moderate $2002-2003$ EN is the only case on the brink of being an extreme event considering the statistics of entropy variations. Thence, from these SOI data it is not possible to consider the transitions between ENSO phases have been entropic extreme events on the whole. Moreover, we have analysed the (mean) entropy variation rate, $\Delta  S / (t_f-t_i)$, for both approaches and could not find any relation between them and the classification of the phases. {\it This result sheds light on the significant contrast between eminently informational based entropy  variations of the indicator, namely $\Delta S (\vec s)$ and $\Delta \tilde S(t_f,t_i)$, and the thermal/heat-related entropy variations in the transition between the El Ni\~no (hotter)$\leftrightarrow$La Ni\~na (colder) ENSO phenomenon phases. Ultimately, this raises the question over the extent of the change in weather conditions when SOI extreme entropic events befall (and they statistically will do so) as the index is used to appraise ENSO states.} Alternatively, this can shift SOI into a minor role on the description of the phenomenon. Regarding that, we plan this study be enlarged to multivariate approaches of ENSO considering not only the temperature dynamics, but also other quantities like the outgoing longwave radiation (anomaly) or precipitation~\cite{an2020,timlin1993,timlin2011,fowler2005,fowler2009,ardanuy1986}. Furthermore, minding the role of entropy in image analysis~\cite{tsai2008} and studies on entropy production and fluctuation relations for surface growth and percolative models~\cite{barato2010}, a SPP methodology can be also implemented on self-organizing maps (som) used to distinguish climate dynamics and its patterns~\cite{gibson2017} by defining $\Delta S  _{\mathrm{som}}$ for sequence sets.
Still, each EN/LN phase can be studied in itself; in this situation, the analysis of (integral) fluctuation relations equivalent to the Speck-Seifert~\cite{seifert2005} and Hatano-Sasa~\cite{sasa2001} are amongst the future problems to be worked at as well.

\smallskip

\begin{acknowledgments}
I acknowledge CNPq, C.B. Naves for software assistance and the comments presented by the Referees from which this work benefitted. 
\end{acknowledgments}

\appendix

\begin{widetext}

\centerline{\large{\sc Supplemental Material}}

\section{Data treatment}

The dataset used in this work consists of the daily records of the SOI spanning the interval between the 157th day of 1991 and 212th day of 2023 and was downloaded from \href{https://www.longpaddock.qld.gov.au/soi/soi-data-files/}{The Long Paddock} website of the Queensland Government (Australia) comprising $11744$ records. 
For the spectral estimation of these data, we have assumed the non-parametric Multi-Taper Method~\cite{thomson1982,percival1993};  tapers~\cite{slepian1978} are the discrete set of eigenfunctions that solve the variational problem over the minimization of leakage outside of a frequency band of half the bandwidth $n \hat f$ where $\hat f = \frac{1}{L \, \tau}$ is the  the Rayleigh frequency ($L$ is the number of elements of the dataset, and $\tau$ the sampling rate) and $n$ is an integer. This method is widely used in the analysis of atmospheric and oceanic data~\cite{ghil2002} among other problems in geophysical signal processing. 

Naturally, the Multi-Taper Method takes into account the application of the $K$ first tapers which provide usefully little spectral-leakage. The literature on the subject~\cite{mann1993} points out the application of $K=3$ for climate records  --- which are typically recorded at sampling rates larger or equal than one month ---, but longer datasets (e.g., by decreasing the sampling rate to a daily basis) admit the utilization of a greater number K of tapers. Taking into consideration previous studies~\cite{ausloos2007} we have considered the spectral analysis with $K=5$.

The background signal/protocol, $\Phi (t)$, is then reconstructed, using the significant frequency peaks, $f_i \equiv T_ i ^{-1}$, shown in Table~\ref{abc},
\begin{equation}
\Phi (t) = \Phi _0 (t) + \sum \Phi _i (t) \cos \left[ \frac{2\, \pi}{T_i} t + \phi _i \right].
\end{equation}
Such significant peaks are determined by the highest peaks of the power spectrum that surpass the $99\%$ confidence interval with respect to Brownian motion established by a $\chi ^2$-distribution with $m = 2\,K$ degrees of freedom. The first (inverse) frequency, $T_1$, is characteristic of the Branstator-Kushnir wave~\cite{branstator1987}, which is a westwards travelling wave with a period of around 23 days); $T_2$, matches the moon rotation; $T_3$ and $T_4$ are within the limits of the Madden and Julian Oscillation, it also concurs with the interseasonal oscillation in the tropical atmosphere~\cite{gasperini2020} though; and $T_6$ is naturally identified with Earth's rotation around the Sun.

\begin{table}[h]  
\caption{Values of the parameters in Eq.~(\ref{protocol}).}  
\centering  
\begin{tabular}{c c c c c c c} 
\hline\hline   
Period (day) && $A_i$ && $B_i$  && $\theta _i $ (rad)
\\   
\hline   
 ----- && $-5.07\pm 0.24$ && $7.9 \times 10^{-4} \pm 4 \times 10^{-5}$ && ----- \\
$T_1 = 24$ && $ 0.493\pm 0.04$ && $ - 3 \times 10^{-5} \pm 6 \times 10^{-5}$ && $6.03 \times 10 ^{-1} \pm 6.0 \times 10 ^{-2}$ \\
$T_2 = 28 $ && $ -0.077\pm 0.03$ && $ - 1 \times 10^{-4}\pm 6 \times 10^{-5}$ && $1.23 \times 10^{-1}\pm 2.7 \times 10 ^{-2}$\\
$T_3 = 36$ && $ 0.041\pm0.04 $ && $- 1.2 \times 10^{-4}\pm 6 \times 10^{-5}$ && $- 1.0 \times 10^{-1}\pm 2.3 \times 10 ^{-2}$\\
$T_4 = 74$ && $ 1.256 \pm 0.33$ && $- 8 \times 10^{-5}\pm 4 \times 10^{-5}$ && $-5.36 \times 10^{-1}\pm 2.4 \times 10 ^{-2}$\\
$T_5 = 102$ && $ -1.979 \pm 0.43$ && $1.4 \times 10^{-4}\pm 5 \times 10^{-5}$ && $-5.44 \times 10^{-2}\pm 1.6 \times 10 ^{-2}$ \\
$T_6 = 365 $ && $ 1.409 \pm 0.31$ && $1.1 \times 10^{-4}\pm 4 \times 10^{-5}$ && $4.92 \times 10^{-1}\pm 9 \times 10 ^{-3}$\\
$T_7 = 2168$ && $ -3.114 \pm 0.33$ && $5 \times 10^{-5}\pm 4 \times 10^{-5}$ && $-3.80 \times 10^{-1}\pm 7 \times 10 ^{-3}$\\
\hline   
\end{tabular} 
\label{abc}
\end{table} 

Both the envelope functions, $\Phi _i (t) $, and phases, $\phi _i $ can be determined from a time-domain inversion of the spectral domain information given by the $K$ complex eigenspectra. However~\cite{mann1996,ghil2002}, in geophysical signal processing this approach is overlooked in favor of a nearly optimal reconstruction obtained by a mean-square minimization of the numerical adjustment of the reconstructed signal with respect to the raw data series leading to,
\vspace{-0.1cm}
\begin{equation}
\Phi (t) = A_0 + B_0 \, t+ \sum _{i=1}^7  \left[ A_i + B_i \,t \right]  \cos  \left[ \frac{2\, \pi}{T_i}t+\theta_i \right].
\label{protocol}
\end{equation}
and finally
%
\begin{equation}
\xi _t = s_t - \Phi (t).
\label{phi+xi}
\end{equation}

The values of the parameters $A_i$, $B_i$, and $\theta _i$ are presented in Table~\ref{abc} as well and hinge on applying the optimization procedure over $10000$ sets of initial conditions of the $\Phi (t) $ parameters (excluding fixed $\{ T \}$.

In Fig.~\ref{fig-signal}, we depict the three signals that we have dealt with: $\{s \}$, $\{\xi \}$, and $\{\Phi \}$.

\begin{figure}[ht!]
\centering
\includegraphics[width = 0.65\linewidth]{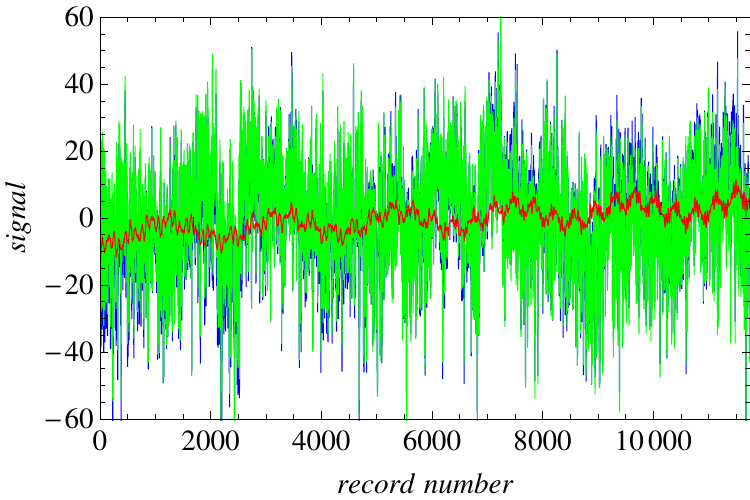}
\caption{The blue line represents the original SOI data, $\{s \}$, the red line is for the background protocol, $\{\Phi \}$, and the green line is the signal $\{\xi \}$.
\label{fig-signal}}
\end{figure}

\section{Computation of the Kramers-Moyal coefficients considering the sampling rate}

In this section, we follow closely the results by Anteneodo \& \surname{Duarte Queir\'os}~\cite{anteneodo2010}. 
Hinging on that work, a time series, ${\boldmath{\xi}} \equiv \{\xi \}$, obtained using a sampling rate $\tau$ from a stochastic differential equation in the It\^o representation,
\begin{equation}
d\xi = D_1(\xi) \,dt + \sqrt{2\, D_2(\xi )} \, dW_t, \qquad \langle W_{t^\prime} \, W_t \rangle = \min (t,t^\prime).
\label{langevin-xi}
\end{equation}
where, 
\begin{equation}
D_1(\xi) = - a \, \xi + b, \qquad D_2(\xi) = \alpha ^2 \xi ^2 + \beta \, \xi + \gamma ^2,
\end{equation}
yields the following effective Kramers-Moyal coefficients,
\begin{equation}
\tilde D_1(\xi, \tau) = - \left( \xi - \frac{b}{a} \right)\frac{1-z}{\tau}, \qquad \tilde D_2(\xi) = \tilde D_2 ^{(2)}(\xi) \, \xi ^2 + \tilde D_2^{(1)}(\xi)\, \xi + \tilde D_2 ^{(0)} (\xi),
\label{D1}
\end{equation}
with,
\begin{eqnarray}
\tilde D_2 ^{(2)}(\xi) &=& \frac{1-2\,z + z^2 \, w}{2\, \tau}, \label{D22}\\
&& \nonumber \\
\tilde D_2^{(1)}(\xi) &=& \frac{1}{\tau} \left[\frac{b+\beta}{a-2\, \alpha ^2}(z-z^2\,w)+\frac{b}{a}(z-1) \right], \label{D21}\\  
&& \nonumber \\
\tilde D_2 ^{(0)} (\xi) &=&   \frac{1}{2\, \tau} \frac{b\, (b+\beta) \,\left[a\, z^2 \, w -2(a -\alpha ^2) \,z + a -2\, \alpha ^2 \right] }{a(a-\alpha ^2)(a-2\, \alpha ^2)} + \frac{\gamma ^2}{a - \alpha ^2} (1-z^2 \, w), \label{D20}
\end{eqnarray}
where $z \equiv \exp[- a\, \tau]$ and $w \equiv \exp[2 \, \alpha ^2 \, \tau]$.

From the numerical adjustments of the empirical Kramers-Moyal moments, $ M_n (\xi)$, so that,
\begin{equation}
\tilde D_n ^\prime (\xi) = \frac{1}{\tau \, n!} M_n (\xi).
\end{equation}
in the panels a) and b) of Fig.~\ref{fig-km} smoothed by the Savitzky-Golay filter~\cite{golay} and owing to the sampling rate $\tau = 1 {\rm day}$ the application of the formulae  we obtain: $a=0.15\pm0.01$, $b=-0.08\pm0.04$, $\alpha ^2 = 0.010\pm 0.002$, $\beta = -0.14\pm 0.03 $ and $\gamma ^2 = 31.65 \pm 0.8$. In the panels c) and d) of Fig.~\ref{fig-km}, we present a straightforward comparison of the third and fourth Kramers-Moyal moments computed from the $\xi $ series obtained from Eq.~(\ref{phi+xi}), $\xi _{{\rm data}}$, and the numerically obtained series, $\xi _{{\rm simulation}}$, assuming Eq.~(\ref{langevin-xi}) with the same number of elements. The agreement between data $\xi $ and the synthetic $\xi $ is $90\% $ statistically significant in the fourth order.

\begin{figure}[ht!]
\centering
\includegraphics[width = 0.45\linewidth]{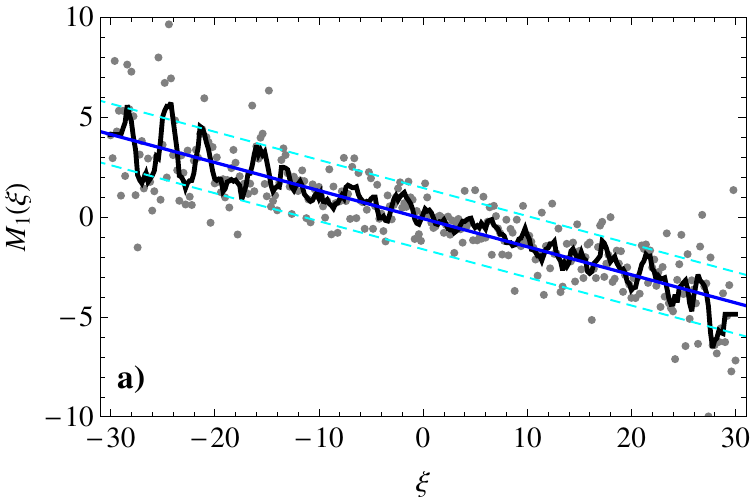}
\includegraphics[width = 0.45\linewidth]{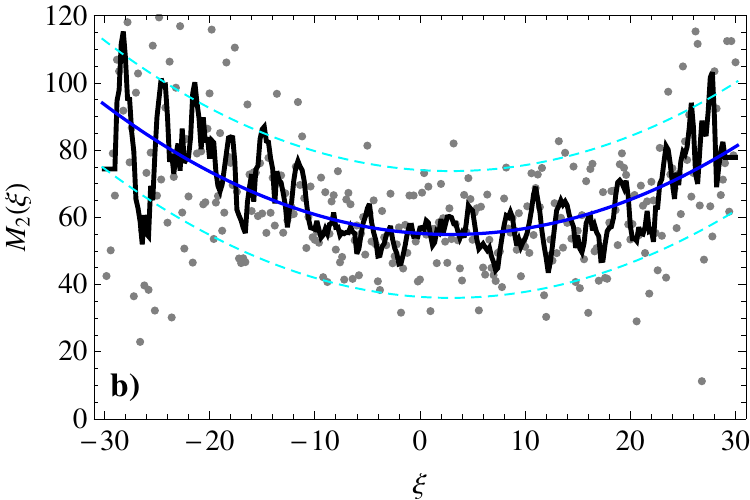}
\includegraphics[width = 0.45\linewidth]{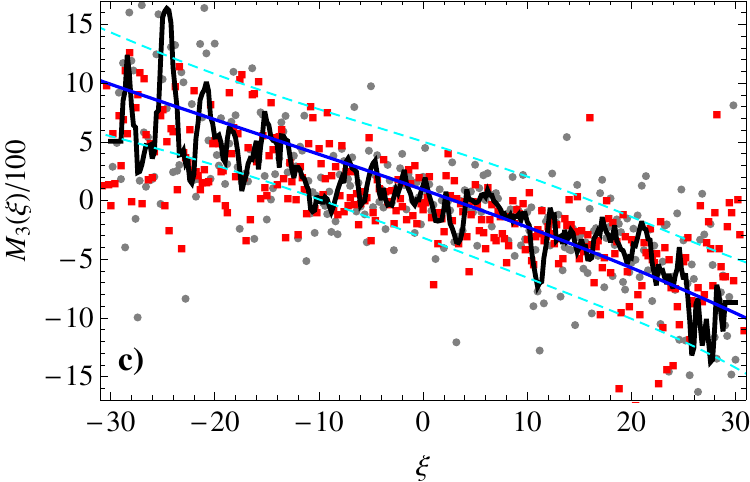}
\includegraphics[width = 0.45\linewidth]{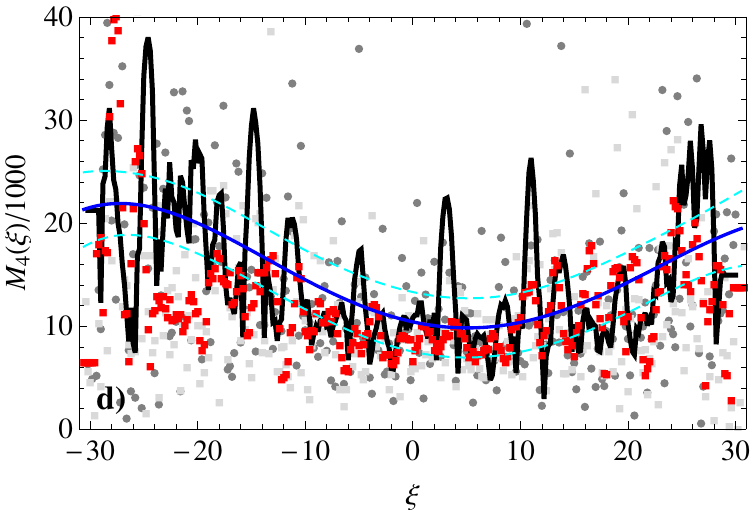}
\caption{First, second, third and fourth empirical Kramers-Moyal moments, $M_n (\xi)$, versus $\xi $ in panels a), b), c), and d), respectively.	The dots are the empirical moments obtained from $\xi _{{\rm data}}$ whereas the squares are the same but for $\xi _{{\rm simulation}}$ instead. The black lines are the smoothed values for $\xi _{{\rm data}}$ using the Savitzky-Golay filter and dashed line is the same for $\xi _{{\rm simulation}}$. The blue lines in a) and b) are the numerical adjustments of the smoothed empirical coefficients whence the parameters of Eq.~(\ref{langevin-xi}) are derived bearing in mind the sampling rate $\tau = 1 {\rm day}$. The cyan dashed lines define the $95\%$ confidence interval. \\
In panels c) and d), we depict a {\it tout court} comparison of the moments yielded by $\xi _{{\rm data}}$ and $\xi _{{\rm simulation}}$, which is significant as stated in this Supplemental Material.
It is worth emphasising that due to the relatively short length of series ($11744$ records), the statistics in the tails turns out poor and strongly distorts the values of the Kramers-Moyal coefficients and its order becomes ever higher.\label{fig-km}}
\end{figure}

\vspace{-0.5cm}
\section{Determination of Eq.~(9) in the Main Text (MT)}

The transformation of a stochastic differential equation in the It\^o interpretation,
\begin{equation}
ds = D_1(s,t) \, dt + \sqrt{2\, D_2(s,t)}\, dW_t,
\label{langevin-soi}
\end{equation}
where is a standard Wiener process ($dW_t = \eta _t \, dt$), into the corresponding stochastic differential equation in the Stratonovich interpretation implies in the introduction of an additional term in the drift~\cite{gardiner1997},
\begin{equation}
ds = \left[ D_1(s,t) - \frac{1}{2}\, D_2(s,t) \frac{\partial}{\partial \, s} \, D_2(s,t) \right]\, dt + \sqrt{2\, D_2(s,t)}\, dW_t,
\label{langevin-soi}
\end{equation}
in which standard derivation rules can be applied. 
To obtain an additive noise process, we consider the change of variables, $y \equiv y(s)$, satisfying~\cite{risken1989},
\begin{equation}
\sqrt{D_2(s,t)} \,\, \frac{\partial \, y}{\partial \, s} = 1
\end{equation}
the solution to which reads~\cite{gradshteyn},
\vspace{-0.4cm}
\begin{equation}
y= \frac{1}{\alpha} \arcsinh \left[ \frac{\alpha}{\lambda} \left(s+\frac{\beta _s}{2 \, \alpha ^2} \right)   \right].
\end{equation}
Carrying out the change of variables using Stratonovich rules,
\begin{equation}
dy = \left\{ \left[ D_1(s,t) - \frac{1}{2}\, D_2(s,t) \frac{\partial}{\partial \, s} \, D_2(s,t) \right]\, dt + \sqrt{2\, D_2(s,t)}\, dW_t \right\} \,\,  \frac{\partial \, y}{\partial \, s},
\label{change-of-variables-strato}
\end{equation}
it yields,
\begin{equation}
dy = \left\{ \left(\frac{b}{\lambda} +\frac{a\, \beta}{2\, \lambda \, \alpha ^2} \right)  \sech [\alpha \, y]  -   \left(\alpha + \frac{a}{\alpha} \right) \tanh [\alpha \, y] \right\} \, dt
+ \sqrt{2} \, dW _t ,
\tag{9-MT}
\label{langevin-y}
\end{equation}

\section{Probabilistic Solution to Eq.~(\ref{langevin-y})}

Recasting Eq.~(\ref{langevin-y}) as,

\begin{equation}
dy = \left\{ \mathcal{A}  \, \sech [\alpha \, y]  -   \mathcal{B} \tanh [\alpha \, y] \right\} \, dt + \sqrt{2} \, dW _t,
\nonumber
\end{equation}

with $\mathcal{A} = \frac{b}{\lambda} +\frac{a\, \beta}{2\, \lambda \, \alpha ^2} $, $\mathcal{B} = \alpha + \frac{a}{\alpha} $.
That dynamics leads to the Fokker-Planck Equation,
\begin{equation}
\frac{\partial \, p(y,t)}{\partial \,t} = - \frac{\partial }{\partial \,y}\left\{ \mathcal{A}  \, \sech [\alpha \, y]  -   \mathcal{B} \tanh [\alpha \, y] \right\} p(y,t) + \frac{\partial ^2 \, p(y,t)}{\partial \,y^2},
\end{equation}
the stationary solution, $\frac{\partial \, p(y,t)}{\partial \,t} = 0$, or
\begin{equation}
\frac{\partial }{\partial \,y} p^*(y)= \left\{ \mathcal{A}  \, \sech [\alpha \, y]  -   \mathcal{B} \, \tanh [\alpha \, y] \right\} p(y,t)
\end{equation}
is equal to,
\begin{equation}
p^*(y)= \frac{1}{\mathcal N}\exp \left[ \frac{2 \, \mathcal{A}}{\alpha } \mathrm{arctanh} \left[ \mathrm{tanh} \left[\frac{\alpha}{2}y \right] \right] \right] \, \cosh \left[  \alpha \, y     \right] ^{-\frac{\mathcal{B}}{\alpha}},
\end{equation}
that leads to the same stationary solution Eq.~(6) in the MT after reverting the transformation of variables and the representation of the noise.

\section{Solution to the SOI Euler-Lagrange equation}

In respect of the time-dependent solution $p(s,t)$, or $p(y,t)$, we need to compute the propagator. As mentioned in the main text, among the several method of getting the propagator we considered that it was best to take the path integral method that we discuss hereinafter.

As stated in the main text, according to the path integral formulation~\cite{wio2013} the conditioned probability distribution, $p(y,t|y_0,t_0)$ is defined as,
\begin{equation}
p(y,t|y_0,t_0) \equiv \int \mathcal{D}y \, \exp \left[ -\frac{1}{4} \mathcal{S}[y(t)]] \right]  \, \det \frac{\delta \eta _t}{\delta y_0},
\label{pathintegral}
\end{equation}
where the of the stochastic action,
%
$\mathcal{S}[y(t)]  \equiv  \int _{t_0} ^t \mathcal{L}(t^\prime) dt^\prime \sim \left[ \eta _t \right]^2,$
%
so that the Onsager-Machlup function reads~\cite{onsager,risken1989},
\begin{equation}
\mathcal{L} =  \left\{ \dot{y} - \left(\frac{b}{\lambda} +\frac{a\, \beta}{2\, \lambda \, \alpha ^2} \right)  \sech [\alpha \, y]  +   \left(\alpha + \frac{a}{\alpha} \right) \tanh [\alpha \, y] \right\} ^2.
\label{onsager-machlup}
\end{equation}

We now split the stochastic trajectory, $y(t)= z(t)+y^*(t)$, into the superposition of fluctuating, $z(t)$, and deterministic paths, $y^*(t)$, with the latter having the same boundary conditions as $y(t)$ whilst it is the optimising solution to the stochastic action,
$$
\delta \, \mathcal{S}[y(t)]] = 0.
$$
The respective Euler-Lagrange equation 
\begin{equation}
\left( \frac{\partial \mathcal{L}}{\partial y}-\frac{d}{dt}\frac{\partial \mathcal{L}}{\partial \dot {y}} \right) \Big\vert _{y=y^*} = 0, 
\label{euler-lagrange}
\end{equation}
explicitly reads,
\begin{eqnarray}
\left\{  \ddot y+ \alpha  \, \dot y \tanh \left[ \alpha  \, y \right] \sech \left[ \alpha  \, y \right] \left(\frac{a
   \beta }{2 \alpha ^2 \lambda }+\frac{b}{\lambda }\right)+\alpha 
   \left(\frac{a}{\alpha }+\alpha \right) \dot y \, \sech ^2\left[ \alpha  \, y \right] \right. - && \nonumber \\
 -\left[\alpha  \tanh \left[ \alpha  \, y \right] \sech \left[ \alpha  \, y \right] \left(\frac{a \beta }{2 \alpha
   ^2 \lambda }+\frac{b}{\lambda }\right)+\alpha  \left(\frac{a}{\alpha }+\alpha
   \right) \sech ^2\left[ \alpha  \, y \right]\right]  \times && \\
  \times  \left. \left[\dot y+\left(\frac{a}{\alpha }+\alpha \right) \tanh (\alpha \, y) - \sech \left[ \alpha  \, y \right] \left(\frac{a \beta }{2 \alpha ^2 \lambda }+\frac{b}{\lambda }\right) \right] \right\} \Big\vert _{y=y^*}&=&0,
  \nonumber
\end{eqnarray}
with boundary conditions,
\begin{equation}
y^*(t) = y, \,\,\,\,y^*(t_0) = y_0.
\end{equation}

Subsequent analysis on the relation between parameters of the problem finally end up in the differential equation,
\begin{equation}
\ddot{y}^*(t) - c_2 \, \left[ y^*(t) \right] ^2 - c_1 \, y^*(t) + c_0 = 0.
\label{euler-lagrange-de}
\end{equation}
with,
\begin{eqnarray}
c_0 &=& \frac{b}{\gamma } (a+\alpha ^2)+ \frac{a\, \beta}{2\, \gamma} \left( 1+ \frac{a}{\alpha ^2} \right), \\
&& \nonumber \\
c_1 &=& a^2 + 2 \, a \, \alpha ^2- (b \, \alpha ^2 + a \, \beta) \frac{b} {\gamma ^2 },\\
&& \nonumber \\
c_2 &=&  \frac{5 a^2 \beta }{4 \gamma }+\frac{5 \alpha ^2 a \beta }{4 \gamma }+\frac{5 \alpha
   ^2 a b}{2 \gamma }.
\end{eqnarray}

Equation~(\ref{euler-lagrange-de}) has still no closed solution and thus we resort to the specific values of the SOI dynamics that imply in $c_2 \approx 7 \times 10^{-5}$, which is 100 times as small as $c_1$. Thence, we neglect the second term on the rhs of Eq.~(\ref{euler-lagrange-de}) that finally give us Eq.~(16) in the Main Text, the solution to which reads,
\begin{equation}
y^*(t ^\prime)= \frac{ c_0}{ c_1 }+ C_1 \, \exp \left[ \sqrt{ c_1 } t ^\prime \right] + C_2 \, \exp \left[-\sqrt{ c_1 } t^\prime \right],
\label{solution-y}
\end{equation}
where the constants $C_1$ and $C_2$ are explicitly obtained bearing in mind the boundary conditions $y^*(t) = y$ and $y^*(t_0) = y_0$. Hereinafter, $k \equiv \sqrt{c_1}$.

\section{Determination and explicit equations for $p \left( s(t)\, | \, s(t_0) \right)$}

With the solution to Eq.~(\ref{euler-lagrange-de}) we establish,
\begin{equation}
p(y,t|y_0,t_0) \equiv    \exp \left[ -\frac{1}{4}\mathcal{S}^* \right]  \,\,\,  \int    \mathcal{D}z \, \exp \left[\mathcal{S}[z(t)] \right]  = \frac{1}{\int \exp \left[ -\frac{1}{4}\mathcal{S}^* \right] \, dy}  \exp \left[ -\frac{1}{4}\mathcal{S}^* \right],
\label{pathintegral}
\end{equation}
%
%
where,
\begin{equation}
\mathcal{S}^*  = \int _{t_0} ^t \mathcal{L}(y^*, \dot y^*; t ^\prime) dt^\prime .
\label{action}
\end{equation}

Plugging all the elements together we get,
\begin{equation}
p \left( s(t)\, | \, s(t_0) ; \Phi \right) = p(y,t|y_0,t_0)\Big\vert _{y=s(y)}  \;\frac{\partial y}{\partial s}.
\label{pdf}
\end{equation}

Although we have carried out hereinafter by employing numerical integration because some integrals in Eq.~(\ref{action}) do not have a closed-form, we can apply follow the previous expansion scheme and present the transition probabilities as follows:
\begin{eqnarray}
p ^F \left( s(t)\, | \, s(t_0) ; \Phi \right) &=& \frac{1}{\mathcal N }\; \exp \left[ -\frac{1}{4} \sum _{i=1} ^6 \Omega _i ^F (s(t),s(t_0);t,t_0,\Phi) \right]  \;\left[  \lambda ^2 + \alpha ^2 \left(s(t)+\frac{\beta _s}{2 \, \alpha ^2} \right)^2 \right]^{-\frac{1}{2}},  \label{pdf-F}   \\
& & \nonumber \\
p ^R \left( s(t_0)\, | \, s(t) ; \Phi \right) &=& \frac{1}{\mathcal N }\; \exp \left[ -\frac{1}{4} \sum _{i=1} ^6 \Omega _i ^R (s(t),s(t_0);t_0,t,\Phi) \right]  \;\left[  \lambda ^2 + \alpha ^2 \left(s(t_0)+\frac{\beta _s}{2 \, \alpha ^2} \right)^2 \right]^{-\frac{1}{2}}.  \label{pdf-R} 
\end{eqnarray}
The arguments of the forward conditioned probability are equal to,
\begin{eqnarray}
\Omega _1 ^F (s(t),s(t_0);t,t_0, \Phi) &=& -\frac{\frac{a^2 \beta }{2 \alpha ^2 \gamma }+\frac{a b}{\gamma }+\frac{a \beta
   }{2 \gamma }+\frac{\alpha ^2 b}{\gamma }}{ \left(1 - e^{
   - 2\, \sqrt{ k } (t - t_0)}\right)^2 \left(a^2+2 \alpha ^2 a-\frac{a b \beta
   }{\gamma ^2}-\frac{\alpha ^2 b^2}{\gamma ^2}\right)^{3/2}} \left[\frac{\left(a \beta +2 \alpha
   ^2 b\right)^2}{4 \alpha ^2 \lambda ^2}-\frac{\alpha ^4 \gamma ^2+a b \beta +\alpha
   ^2 b^2}{\gamma ^2}\right] \times \nonumber \\
   && \times  \left(\frac{\arcsinh\left[\frac{\alpha 
   \left(\frac{\beta }{2 \alpha ^2}+s-\Phi (t)\right)}{\lambda }\right]}{\alpha
   }+\frac{\arcsinh\left[\frac{\alpha  \left(\frac{\beta }{2 \alpha
   ^2}+ s_0 -\Phi ( t_0 )\right)}{\lambda }\right]}{\alpha }\right) + \nonumber \\
  && 
  +\frac{\alpha ^4+2
   a^2+4 a \alpha ^2-\frac{b \left(a \beta +\alpha ^2 b\right)}{\gamma
   ^2}-\frac{\left(a \beta +2 \alpha ^2 b\right)^2}{4 \alpha ^2 \lambda ^2}}
   {2 \left(1 -e^{ - 2 \sqrt{ k }
   ( t -t_0)} \right)^2 \sqrt{a^2+2 \alpha ^2 a-\frac{a b \beta }{\gamma
   ^2}-\frac{\alpha ^2 b^2}{\gamma ^2}}} \times \nonumber \\
   && \times \left(\frac{\arcsinh\left[\frac{\alpha  \left(\frac{\beta }{2 \alpha ^2}+s-\Phi
   (t)\right)}{\lambda }\right]^2}{\alpha ^2}+\frac{\arcsinh\left[\frac{\alpha 
   \left(\frac{\beta }{2 \alpha ^2}+ s_0 -\Phi ( t_0 )\right)}{\lambda
   }\right]^2}{\alpha ^2}\right) - \\
   && -\frac{a \beta +2 \alpha ^2 b}{6\, 
   \alpha ^3 \lambda }
   \left(\arcsinh ^3 \left[\frac{\alpha  \left(\frac{\beta }{2
   \alpha ^2}+s-\Phi (t)\right)}{\lambda }\right]-\arcsinh ^3
   \left[\frac{\alpha  \left(\frac{\beta }{2 \alpha ^2}+ s_0 -\Phi
   ( t_0 )\right)}{\lambda }\right]\right) - \nonumber \\
   && -\frac{a \beta +2 \alpha ^2 b}{
   \alpha ^3 \lambda }
   \left( \arcsinh\left[\frac{\alpha  \left(\frac{\beta }{2 \alpha ^2}+s-\Phi (t)\right)}{\lambda
   }\right]-\arcsinh\left[\frac{\alpha  \left(\frac{\beta }{2
   \alpha ^2}+ s_0 -\Phi ( t_0 )\right)}{\lambda }\right]\right) + \nonumber \\
   && +\frac{\alpha ^2+a}{\alpha} \left(\arcsinh ^2
   \left[\frac{\alpha  \left(\frac{\beta }{2 \alpha ^2}+s-\Phi (t)\right)}{\lambda
   }\right]-\arcsinh ^2 \left[\frac{\alpha  \left(\frac{\beta }{2 \alpha
   ^2}+ s_0 -\Phi ( t_0 )\right)}{\lambda }\right]\right), \nonumber
\end{eqnarray}

\begin{eqnarray}
\Omega _2 ^F (s(t),s(t_0);t,t_0, \Phi) &=& \frac{ e^{- \sqrt{ k } (t - t_0)} \left(\alpha ^2+a\right)
   \left(\gamma ^2 \left(a \beta +2 \alpha ^2 b\right)^3-4 \alpha ^2 \lambda ^2 \left(a
   \beta +2 \alpha ^2 b\right) \left(\alpha ^4 \gamma ^2+a b \beta +\alpha ^2
   b^2\right)\right) }{\left(1 - e^{ - 2\, \sqrt{ k } (t - t_0)}\right)^2 8 \alpha ^4 \gamma ^3 \lambda
   ^2 \left(a^2+2 \alpha ^2
    a-\frac{a b \beta }{\gamma ^2}-\frac{\alpha ^2 b^2}{\gamma ^2}\right) \sqrt{a^2+2 \alpha 
    ^2a-\frac{a b   \beta }{\gamma ^2}-\frac{\alpha ^2 b^2}{\gamma ^2}}} \nonumber \\
   && - \frac{2
   e^{- \sqrt{ k } (t - t_0)} \left(\alpha ^4+2 a^2+4 a \alpha ^2-\frac{b
   \left(a \beta +\alpha ^2 b\right)}{\gamma ^2}-\frac{\left(a \beta +2 \alpha ^2
   b\right)^2}{4 \alpha ^2 \lambda ^2}\right)}{ \left(1 - e^{- 2 \sqrt{ k }
   ( t -t_0)}\right) \alpha ^2 \sqrt{a^2+2 \alpha ^2 a-\frac{a b
   \beta }{\gamma ^2}-\frac{\alpha ^2 b^2}{\gamma ^2}}}  \times \nonumber \\
   && \times \arcsinh \left(\frac{\alpha 
   \left(\frac{\beta }{2 \alpha ^2}+s-\Phi (t)\right)}{\lambda }\right) \arcsinh
   \left(\frac{\alpha  \left(\frac{\beta }{2 \alpha ^2}+ s_0 -\Phi
   ( t_0 )\right)}{\lambda }\right) \nonumber \\
   && -\frac{e^{- 4\, \sqrt{ k } (t - t_0)} \left(\alpha
   ^4+2 a^2+4 a \alpha ^2-\frac{b \left(a \beta +\alpha ^2 b\right)}{\gamma
   ^2}-\frac{\left(a \beta +2 \alpha ^2 b\right)^2}{4 \alpha ^2 \lambda ^2}\right)}
   {2  \alpha ^2 \left(1- e^{- 2 \sqrt{ k }
   ( t -t_0)}\right)^2 \sqrt{a^2+2 \alpha ^2 a-\frac{a b
   \beta }{\gamma ^2}-\frac{\alpha ^2 b^2}{\gamma ^2}}} \times \\ 
   && \times \left(\arcsinh ^2 \left(\frac{\alpha  \left(\frac{\beta }{2 \alpha ^2}+s-\Phi
   (t)\right)}{\lambda }\right)+\arcsinh ^2 \left(\frac{\alpha 
   \left(\frac{\beta }{2 \alpha ^2}+ s_0 -\Phi ( t_0 )\right)}{\lambda
   }\right)\right) \nonumber \\
   &&+ \tanh \left( - \frac{1}{2} \sqrt{ k } (t - t_0)\right) \frac{\left(\alpha ^2+a\right) \left(a \beta +2 
   \alpha  ^2 b\right)}
   {\alpha ^3 \lambda \sqrt{a^2+2 \alpha ^2 a-\frac{a b
   \beta }{\gamma ^2}-\frac{\alpha ^2 b^2}{\gamma ^2}}}  \times \nonumber \\
   &&\times \left(\arcsinh \left(\frac{\alpha  \left(\frac{\beta }{2 \alpha ^2}+s-\Phi
   (t)\right)}{\lambda }\right)+ \arcsinh \left(\frac{\alpha 
   \left(\frac{\beta }{2 \alpha ^2}+ s_0 -\Phi ( t_0 )\right)}{\lambda
   }\right)\right) \nonumber,
\end{eqnarray}

\begin{eqnarray}
\Omega _3 ^F (s(t),s(t_0);t,t_0, \Phi) &=&  (t- t_0 ) \left(a^2+2 \alpha ^2 a+\alpha ^2 \left(\frac{\left(\frac{a \beta }{\alpha ^2}+2 b\right)^2}{4 \lambda
   ^2}-\left(\frac{a}{\alpha }+\alpha \right)^2\right)-\frac{a b \beta }{\gamma ^2}-\frac{\alpha ^2 b^2}{\gamma ^2}\right) \times \nonumber \\
&&   \times \left\{  \sech^2\left(\frac{1}{2} \sqrt{ k } (t- t_0 )\right) \frac{ \left(\frac{a^2 \beta }{2 \alpha ^2 \gamma
   }+\frac{a b}{\gamma }+\frac{a \beta }{2 \gamma }+\frac{\alpha ^2 b}{\gamma }\right)}{2 \alpha \left(a^2+2 \alpha ^2
   a-\frac{a b \beta }{\gamma ^2}-\frac{\alpha ^2 b^2}{\gamma ^2}\right)} \right. \times \nonumber \\
   && \times 
   \left[\arcsinh \left(\frac{\alpha 
   \left(\frac{\beta }{2 \alpha ^2}+s-\Phi (t)\right)}{\lambda }\right)+\arcsinh \left(\frac{\alpha 
   \left(\frac{\beta }{2 \alpha ^2}+ s_0 -\Phi ( t_0 )\right)}{\lambda }\right)\right] + \\
   && +\frac{1}{2 \alpha ^2} \csch^2\left(\sqrt{ k }
   (t- t_0 )\right) \left[\arcsinh ^2\left(\frac{\alpha  \left(\frac{\beta }{2 \alpha ^2}+s-\Phi (t)\right)}{\lambda
   }\right)+\arcsinh ^2\left(\frac{\alpha  \left(\frac{\beta }{2 \alpha ^2}+ s_0 -\Phi
   ( t_0 )\right)}{\lambda }\right) \right] - \nonumber \\ 
   && -  \left. \coth \left(- \sqrt{ k } ( t -t_0)\right)
   \csch\left(- \sqrt{ k } ( t -t_ 0)\right) \frac{ \arcsinh \left(\frac{\alpha  \left(\frac{\beta }{2 \alpha ^2}+s-\Phi
   (t)\right)}{\lambda }\right) \arcsinh \left(\frac{\alpha  \left(\frac{\beta }{2 \alpha ^2}+ s_0 -\Phi
   ( t_0 )\right)}{\lambda }\right)}{\alpha ^2} \right\} \nonumber,
\end{eqnarray}

\begin{eqnarray}
\Omega _4 ^F (s(t),s(t_0);t,t_0, \Phi) &=& -\frac{\tanh
   \left(- \frac{1}{2} \sqrt{ k } (t - t_0)\right)}{8 \alpha ^4 \gamma  \lambda ^2 \left(a^2+2 \alpha ^2 a-\frac{a b
   \beta }{\gamma ^2}-\frac{\alpha ^2 b^2}{\gamma ^2}\right)^{3/2} \left(a \gamma ^2
   \left(2 \alpha ^2+a\right)-a b \beta +\alpha ^2 \left(-b^2\right)\right)} \times  \nonumber \\
   && \left(\alpha ^2+a\right) \left(a \beta +2 \alpha ^2 b\right)  \left(\frac{a^2 \beta }{2
   \alpha ^2 \gamma }+\frac{a b}{\gamma }+\frac{a \beta }{2 \gamma }+\frac{\alpha ^2
   b}{\gamma }\right) \times  \\
   && \left[\gamma ^2 \left(-12 \alpha ^6 \lambda ^2+a^2 \left(8 \alpha
   ^2 \lambda  (2 \gamma -\lambda )+3 \beta ^2\right)+16 a \alpha ^4 \lambda  (2 \gamma
   -\lambda )\right) \right. + \nonumber \\ 
   && \left. + 4 a \alpha ^2 b \beta  \left(3 \gamma ^2-4 \gamma  \lambda
   -\lambda ^2\right)+4 \alpha ^4 b^2 \left(3 \gamma ^2-4 \gamma  \lambda -\lambda
   ^2\right)\right] \nonumber,
\end{eqnarray}

\begin{eqnarray}
\Omega _5 ^F (s(t),s(t_0);t,t_0, \Phi) &=& (t- t_0 )\frac{\gamma ^2 \left(\alpha ^2+a\right)^2  \left(a \beta +2 \alpha ^2
   b\right)^2 \left(4 \alpha ^2 \lambda ^2
   \left(\alpha ^2+a\right)^2 - a \beta +2 \alpha ^2 b\right)^2}{16 \alpha ^6 \lambda ^2 \left(1- e^{
   - 2 \sqrt{ k } (t - t_0)}\right)^2 \left(a b \beta +\alpha ^2 b^2\right)^2-a 
   \gamma ^2 \left(2 \alpha ^2+a\right)}+ \nonumber \\
   && (t- t_0 ) \frac{\gamma  \left(\alpha
   ^2+a\right)^2  \left(a \beta +2 \alpha ^2 b\right)^2}{2 \alpha ^4
   \lambda  \left(-a \gamma ^2 \left(2 \alpha ^2+a\right)+a b \beta +\alpha ^2
   b^2\right)}+(t- t_0 ) \frac{ \left(a \beta +2 \alpha ^2 b\right)^2}{4 \alpha ^4
   \lambda ^2},
\end{eqnarray}

\begin{eqnarray}
\Omega _6 ^F (s(t),s(t_0);t,t_0, \Phi) &=& e^{- \sqrt{ k } (t - t_0)} \left(\frac{a^2 \beta }{2 \alpha ^2 \gamma }+\frac{a
   b}{\gamma }+\frac{a \beta }{2 \gamma }+\frac{\alpha ^2 b}{\gamma }\right)^2
   \left[\frac{2}{\left(1+ e^{- \sqrt{ k } (t - t_0)}\right)^2 \left(\frac{a b \beta }{\gamma ^2}+
   \frac{\alpha ^2 b^2}{\gamma ^2}-a^2-2 \alpha ^2 a\right)}- \right. \\
   && \left. - \frac{\gamma ^4 \left(2 -6 e^{- 
   \sqrt{ k } (t - t_0)}+2 e^{ - 2\, \sqrt{ k } (t - t_0)}+e^{- 3\, \sqrt{ k } (t - t_0)}\right)
   \left(\left(a \beta +2 \alpha ^2 b\right)^2-4 \alpha ^2 \lambda ^2 \left(\alpha
   ^2+a\right)^2\right)}{4 \alpha ^2 \lambda ^2 \left(1-e^{- 2 \sqrt{ k }
   ( t -t_0)}\right)^2 \left(a b \beta
   +\alpha ^2 b^2-a \gamma ^2 \left(2 \alpha ^2+a\right)\right)^2}\right] \nonumber .
\end{eqnarray}

Concerning the arguments of the reverse conditioned probability they are equal to,
\begin{eqnarray}
\Omega _1 ^R (s(t_0),s(t);t_0,t, \Phi) &=& -\frac{\frac{a^2 \beta }{2 \alpha ^2 \gamma }+\frac{a b}{\gamma }+\frac{a \beta
   }{2 \gamma }+\frac{\alpha ^2 b}{\gamma }}{ \left(1 - e^{
   - 2\, \sqrt{ k } | t - t_0 |}\right)^2 \left(a^2+2 \alpha ^2 a-\frac{a b \beta
   }{\gamma ^2}-\frac{\alpha ^2 b^2}{\gamma ^2}\right)^{3/2}} \left[\frac{\left(a \beta +2 \alpha
   ^2 b\right)^2}{4 \alpha ^2 \lambda ^2}-\frac{\alpha ^4 \gamma ^2+a b \beta +\alpha
   ^2 b^2}{\gamma ^2}\right] \times \nonumber \\
   && \times  \left(\frac{\arcsinh\left[\frac{\alpha 
   \left(\frac{\beta }{2 \alpha ^2}+s_0-\Phi ( t_0 )\right)}{\lambda }\right]}{\alpha
   }+\frac{\arcsinh\left[\frac{\alpha  \left(\frac{\beta }{2 \alpha
   ^2}+ s -\Phi ( t )\right)}{\lambda }\right]}{\alpha }\right) + \nonumber \\
  && 
  +\frac{\alpha ^4+2
   a^2+4 a \alpha ^2-\frac{b \left(a \beta +\alpha ^2 b\right)}{\gamma
   ^2}-\frac{\left(a \beta +2 \alpha ^2 b\right)^2}{4 \alpha ^2 \lambda ^2}}
   {2 \left(1 -e^{ - 2 \sqrt{ k }
   ( t -t_0)} \right)^2 \sqrt{a^2+2 \alpha ^2 a-\frac{a b \beta }{\gamma
   ^2}-\frac{\alpha ^2 b^2}{\gamma ^2}}} \times \nonumber \\
   && \times \left(\frac{\arcsinh\left[\frac{\alpha  \left(\frac{\beta }{2 \alpha ^2}+s_0-\Phi
   ( t_0 )\right)}{\lambda }\right]^2}{\alpha ^2}+\frac{\arcsinh\left[\frac{\alpha 
   \left(\frac{\beta }{2 \alpha ^2}+ s -\Phi ( t )\right)}{\lambda
   }\right]^2}{\alpha ^2}\right) - \\
   && -\frac{a \beta +2 \alpha ^2 b}{6\, 
   \alpha ^3 \lambda }
   \left(\arcsinh ^3 \left[\frac{\alpha  \left(\frac{\beta }{2
   \alpha ^2}+s_0-\Phi ( t_0 )\right)}{\lambda }\right]-\arcsinh ^3
   \left[\frac{\alpha  \left(\frac{\beta }{2 \alpha ^2}+ s -\Phi
   ( t )\right)}{\lambda }\right]\right) - \nonumber \\
   && -\frac{a \beta +2 \alpha ^2 b}{
   \alpha ^3 \lambda }
   \left( \arcsinh\left[\frac{\alpha  \left(\frac{\beta }{2 \alpha ^2}+s_0-\Phi ( t_0 )\right)}{\lambda
   }\right]-\arcsinh\left[\frac{\alpha  \left(\frac{\beta }{2
   \alpha ^2}+ s -\Phi ( t )\right)}{\lambda }\right]\right) + \nonumber \\
   && +\frac{\alpha ^2+a}{\alpha} \left(\arcsinh ^2
   \left[\frac{\alpha  \left(\frac{\beta }{2 \alpha ^2}+s_0-\Phi ( t_0 )\right)}{\lambda
   }\right]-\arcsinh ^2 \left[\frac{\alpha  \left(\frac{\beta }{2 \alpha
   ^2}+ s -\Phi ( t )\right)}{\lambda }\right]\right), \nonumber
\end{eqnarray}

\begin{eqnarray}
\Omega _2 ^R (s(t_0),s(t);t_0,t, \Phi) &=& \frac{ e^{- \sqrt{ k } | t - t_0 |} \left(\alpha ^2+a\right)
   \left(\gamma ^2 \left(a \beta +2 \alpha ^2 b\right)^3-4 \alpha ^2 \lambda ^2 \left(a
   \beta +2 \alpha ^2 b\right) \left(\alpha ^4 \gamma ^2+a b \beta +\alpha ^2
   b^2\right)\right) }{\left(1 - e^{ - 2\, \sqrt{ k } | t - t_0 |}\right)^2 8 \alpha ^4 \gamma ^3 \lambda
   ^2 \left(a^2+2 \alpha ^2
    a-\frac{a b \beta }{\gamma ^2}-\frac{\alpha ^2 b^2}{\gamma ^2}\right) \sqrt{a^2+2 \alpha 
    ^2a-\frac{a b   \beta }{\gamma ^2}-\frac{\alpha ^2 b^2}{\gamma ^2}}} \nonumber \\
   && - \frac{2
   e^{- \sqrt{ k } | t - t_0 |} \left(\alpha ^4+2 a^2+4 a \alpha ^2-\frac{b
   \left(a \beta +\alpha ^2 b\right)}{\gamma ^2}-\frac{\left(a \beta +2 \alpha ^2
   b\right)^2}{4 \alpha ^2 \lambda ^2}\right)}{ \left(1 - e^{- 2 \sqrt{ k }
   ( t -t_0)}\right) \alpha ^2 \sqrt{a^2+2 \alpha ^2 a-\frac{a b
   \beta }{\gamma ^2}-\frac{\alpha ^2 b^2}{\gamma ^2}}}  \times \nonumber \\
   && \times \arcsinh \left(\frac{\alpha 
   \left(\frac{\beta }{2 \alpha ^2}+s_0-\Phi ( t_0 )\right)}{\lambda }\right) \arcsinh
   \left(\frac{\alpha  \left(\frac{\beta }{2 \alpha ^2}+ s -\Phi
   ( t )\right)}{\lambda }\right) \nonumber \\
   && -\frac{e^{- 4\, \sqrt{ k } | t - t_0 |} \left(\alpha
   ^4+2 a^2+4 a \alpha ^2-\frac{b \left(a \beta +\alpha ^2 b\right)}{\gamma
   ^2}-\frac{\left(a \beta +2 \alpha ^2 b\right)^2}{4 \alpha ^2 \lambda ^2}\right)}
   {2  \alpha ^2 \left(1- e^{- 2 \sqrt{ k }
   ( t -t_0)}\right)^2 \sqrt{a^2+2 \alpha ^2 a-\frac{a b
   \beta }{\gamma ^2}-\frac{\alpha ^2 b^2}{\gamma ^2}}} \times \\ 
   && \times \left(\arcsinh ^2 \left(\frac{\alpha  \left(\frac{\beta }{2 \alpha ^2}+s_0-\Phi
   ( t_0 )\right)}{\lambda }\right)+\arcsinh ^2 \left(\frac{\alpha 
   \left(\frac{\beta }{2 \alpha ^2}+ s -\Phi ( t )\right)}{\lambda
   }\right)\right) \nonumber \\
   &&+ \tanh \left( - \frac{1}{2} \sqrt{ k } | t - t_0 |\right) \frac{\left(\alpha ^2+a\right) \left(a \beta +2 
   \alpha  ^2 b\right)}
   {\alpha ^3 \lambda \sqrt{a^2+2 \alpha ^2 a-\frac{a b
   \beta }{\gamma ^2}-\frac{\alpha ^2 b^2}{\gamma ^2}}}  \times \nonumber \\
   &&\times \left(\arcsinh \left(\frac{\alpha  \left(\frac{\beta }{2 \alpha ^2}+s_0-\Phi
   ( t_0 )\right)}{\lambda }\right)+ \arcsinh \left(\frac{\alpha 
   \left(\frac{\beta }{2 \alpha ^2}+ s -\Phi ( t )\right)}{\lambda
   }\right)\right) \nonumber,
\end{eqnarray}

\begin{eqnarray}
\Omega _3 ^R (s(t_0),s(t);t_0,t,, \Phi) &=&  |t- t_0 | \left(a^2+2 \alpha ^2 a+\alpha ^2 \left(\frac{\left(\frac{a \beta }{\alpha ^2}+2 b\right)^2}{4 \lambda
   ^2}-\left(\frac{a}{\alpha }+\alpha \right)^2\right)-\frac{a b \beta }{\gamma ^2}-\frac{\alpha ^2 b^2}{\gamma ^2}\right) \times \nonumber \\
&&   \times \left\{  \sech^2\left(\frac{1}{2} \sqrt{ k } |t- t_0 |\right) \frac{ \left(\frac{a^2 \beta }{2 \alpha ^2 \gamma
   }+\frac{a b}{\gamma }+\frac{a \beta }{2 \gamma }+\frac{\alpha ^2 b}{\gamma }\right)}{2 \alpha \left(a^2+2 \alpha ^2
   a-\frac{a b \beta }{\gamma ^2}-\frac{\alpha ^2 b^2}{\gamma ^2}\right)} \right. \times \nonumber \\
   && \times 
   \left[\arcsinh \left(\frac{\alpha 
   \left(\frac{\beta }{2 \alpha ^2}+s_0-\Phi ( t_0 )\right)}{\lambda }\right)+\arcsinh \left(\frac{\alpha 
   \left(\frac{\beta }{2 \alpha ^2}+ s -\Phi ( t )\right)}{\lambda }\right)\right] + \\
   && +\frac{1}{2 \alpha ^2} \csch^2\left(\sqrt{ k }
   |t- t_0 |\right) \left[\arcsinh ^2\left(\frac{\alpha  \left(\frac{\beta }{2 \alpha ^2}+s_0-\Phi ( t_0 )\right)}{\lambda
   }\right)+\arcsinh ^2\left(\frac{\alpha  \left(\frac{\beta }{2 \alpha ^2}+ s -\Phi
   ( t )\right)}{\lambda }\right) \right] - \nonumber \\ 
   && -  \left. \coth \left(- \sqrt{ k } ( t -t_0)\right)
   \csch\left(- \sqrt{ k } ( t -t_ 0)\right) \frac{ \arcsinh \left(\frac{\alpha  \left(\frac{\beta }{2 \alpha ^2}+s_0-\Phi
   ( t_0 )\right)}{\lambda }\right) \arcsinh \left(\frac{\alpha  \left(\frac{\beta }{2 \alpha ^2}+ s -\Phi
   ( t )\right)}{\lambda }\right)}{\alpha ^2} \right\} \nonumber,
\end{eqnarray}

\begin{eqnarray}
\Omega _4 ^R (s(t_0),s(t);t_0,t,, \Phi) &=& -\frac{\tanh
   \left(- \frac{1}{2} \sqrt{ k } | t - t_0 |\right)}{8 \alpha ^4 \gamma  \lambda ^2 \left(a^2+2 \alpha ^2 a-\frac{a b
   \beta }{\gamma ^2}-\frac{\alpha ^2 b^2}{\gamma ^2}\right)^{3/2} \left(a \gamma ^2
   \left(2 \alpha ^2+a\right)-a b \beta +\alpha ^2 \left(-b^2\right)\right)} \times  \nonumber \\
   && \left(\alpha ^2+a\right) \left(a \beta +2 \alpha ^2 b\right)  \left(\frac{a^2 \beta }{2
   \alpha ^2 \gamma }+\frac{a b}{\gamma }+\frac{a \beta }{2 \gamma }+\frac{\alpha ^2
   b}{\gamma }\right) \times  \\
   && \left[\gamma ^2 \left(-12 \alpha ^6 \lambda ^2+a^2 \left(8 \alpha
   ^2 \lambda  (2 \gamma -\lambda )+3 \beta ^2\right)+16 a \alpha ^4 \lambda  (2 \gamma
   -\lambda )\right) \right. + \nonumber \\ 
   && \left. + 4 a \alpha ^2 b \beta  \left(3 \gamma ^2-4 \gamma  \lambda
   -\lambda ^2\right)+4 \alpha ^4 b^2 \left(3 \gamma ^2-4 \gamma  \lambda -\lambda
   ^2\right)\right] \nonumber,
\end{eqnarray}

\begin{eqnarray}
\Omega _5 ^R (s(t_0),s( t );t_0,t,, \Phi) &=& |t- t_0 |\frac{\gamma ^2 \left(\alpha ^2+a\right)^2  \left(a \beta +2 \alpha ^2
   b\right)^2 \left(4 \alpha ^2 \lambda ^2
   \left(\alpha ^2+a\right)^2 - a \beta +2 \alpha ^2 b\right)^2}{16 \alpha ^6 \lambda ^2 \left(1- e^{
   - 2 \sqrt{ k } | t - t_0 |}\right)^2 \left(a b \beta +\alpha ^2 b^2\right)^2-a 
   \gamma ^2 \left(2 \alpha ^2+a\right)}+ \nonumber \\
   && |t- t_0 | \frac{\gamma  \left(\alpha
   ^2+a\right)^2  \left(a \beta +2 \alpha ^2 b\right)^2}{2 \alpha ^4
   \lambda  \left(-a \gamma ^2 \left(2 \alpha ^2+a\right)+a b \beta +\alpha ^2
   b^2\right)}+|t- t_0 | \frac{ \left(a \beta +2 \alpha ^2 b\right)^2}{4 \alpha ^4
   \lambda ^2},
\end{eqnarray}

\begin{eqnarray}
\Omega _6 ^R (s(t_0),s( t );t_0,t,, \Phi) &=& e^{- \sqrt{ k } | t - t_0 |} \left(\frac{a^2 \beta }{2 \alpha ^2 \gamma }+\frac{a
   b}{\gamma }+\frac{a \beta }{2 \gamma }+\frac{\alpha ^2 b}{\gamma }\right)^2
   \left[\frac{2}{\left(1+ e^{- \sqrt{ k } | t - t_0 |}\right)^2 \left(\frac{a b \beta }{\gamma ^2}+
   \frac{\alpha ^2 b^2}{\gamma ^2}-a^2-2 \alpha ^2 a\right)}- \right. \\
   && \left. - \frac{\gamma ^4 \left(2 -6 e^{- 
   \sqrt{ k } | t - t_0 |}+2 e^{ - 2\, \sqrt{ k } | t - t_0 |}+e^{- 3\, \sqrt{ k } | t - t_0 |}\right)
   \left(\left(a \beta +2 \alpha ^2 b\right)^2-4 \alpha ^2 \lambda ^2 \left(\alpha
   ^2+a\right)^2\right)}{4 \alpha ^2 \lambda ^2 \left(1-e^{- 2 \sqrt{ k }
   ( t -t_0)}\right)^2 \left(a b \beta
   +\alpha ^2 b^2-a \gamma ^2 \left(2 \alpha ^2+a\right)\right)^2}\right] \nonumber .
\end{eqnarray}

In Fig.~\ref{fig-jpdf}, we depict the joint probability distribution function,
\begin{equation}
f[s(t),s(t_0);t,t_0;\Phi(t),\Phi(t_0)]=p^F(s(t)\,|\,s(t_0);\Phi(t),\Phi(t_0)) \, p_{t_0}^*(s(t_0);\Phi(t_0)),
\end{equation}
for $t_0=1 \, {\rm day}$ (the start of our daily SOI dataset) and $ t= 453 \, {\rm days}$ (the start of the first SOI transition in our dataset) as an illustration.

\begin{figure}[ht!]
\centering
\includegraphics[width = 0.5\linewidth]{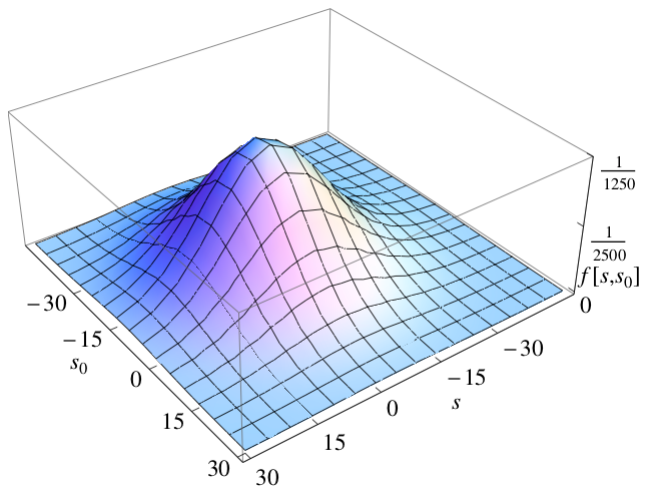}
\caption{Joint (forward) probability density function $f[s(t),s(t_0);t,t_0;\Phi(t)$  for $t_0=1 \, {\rm day}$ and $ t= 453 \, {\rm days}$ .
\label{fig-jpdf}}
\end{figure}

\newpage
\section{Likelihood of the entropic events}

In Fig.~\ref{fig-cdf}, we depict the cumulative distribution functions,
\begin{equation}
P(\Delta S) \equiv \int _{-\infty} ^{\Delta S} p(x) \, dx,
\end{equation}
for the two cases we have analysed, namely $\Delta S (\vec s)$ and $\Delta \tilde S (t_f,t_i)$.  On each curve, we have marked  the realised value of the entropy variation according to the values presented in Tables~II and~III in the MT.

\begin{figure}[ht!]
\centering
\includegraphics[width = 0.35\linewidth]{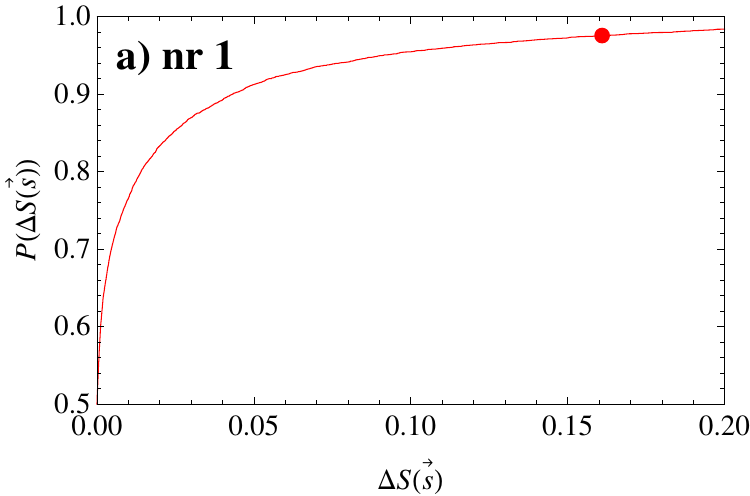}
\includegraphics[width = 0.35\linewidth]{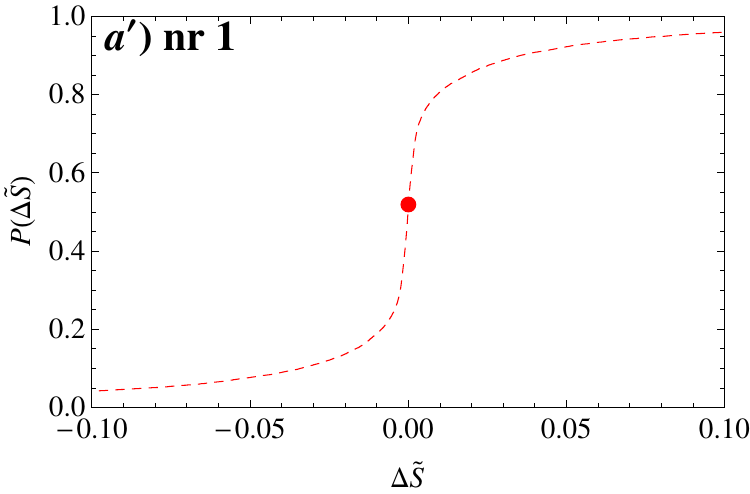}
\end{figure}
\begin{figure}[ht!]
\includegraphics[width = 0.35\linewidth]{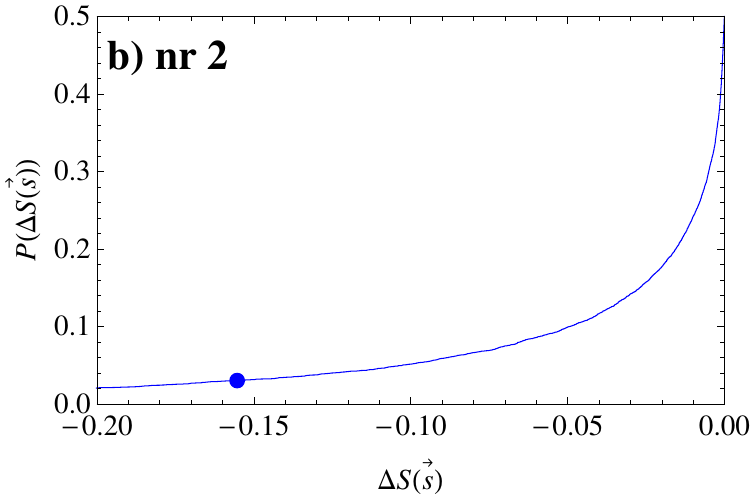}
\includegraphics[width = 0.35\linewidth]{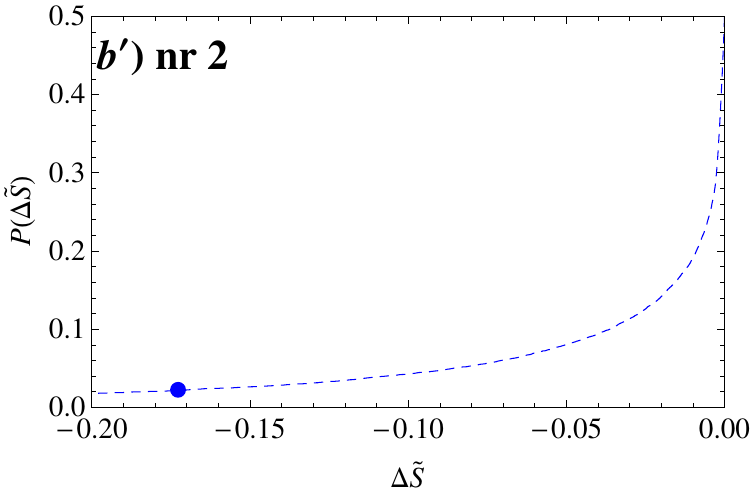}
\end{figure}
\begin{figure}[ht!]
\includegraphics[width = 0.35\linewidth]{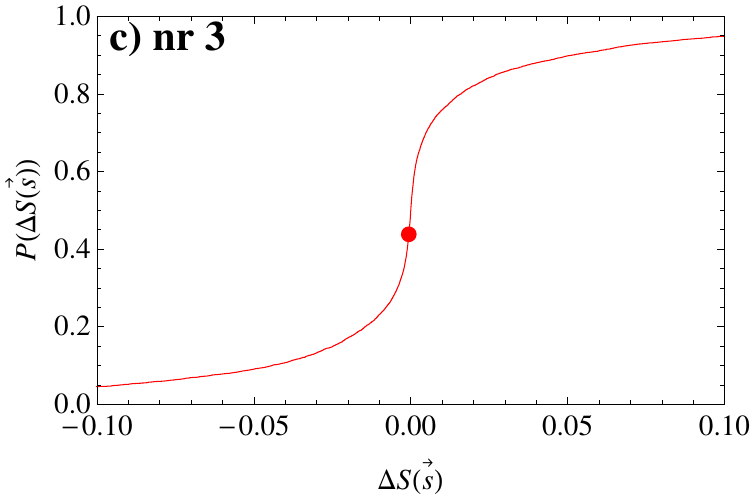}
\includegraphics[width = 0.35\linewidth]{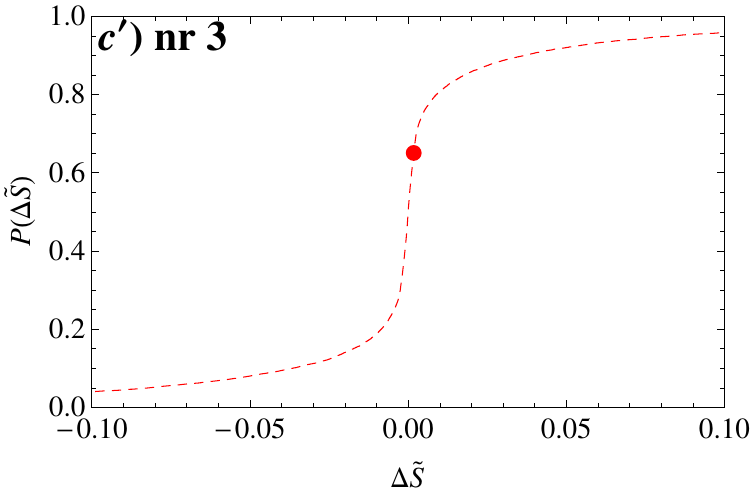}
\end{figure}
\begin{figure}[ht!]
\includegraphics[width = 0.35\linewidth]{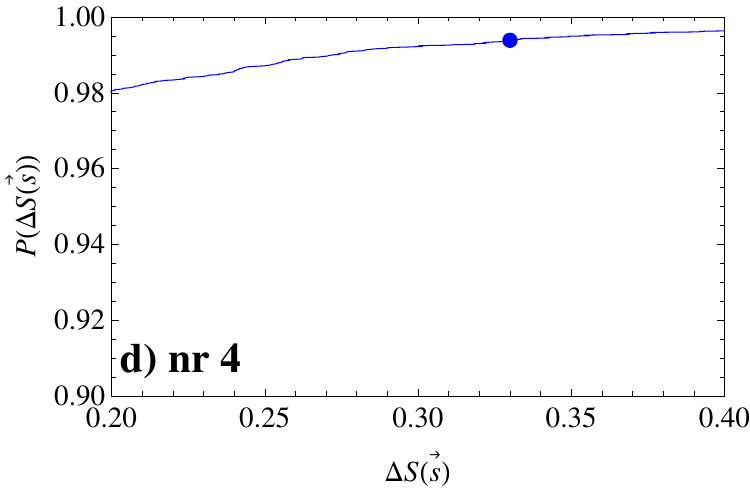}
\includegraphics[width = 0.35\linewidth]{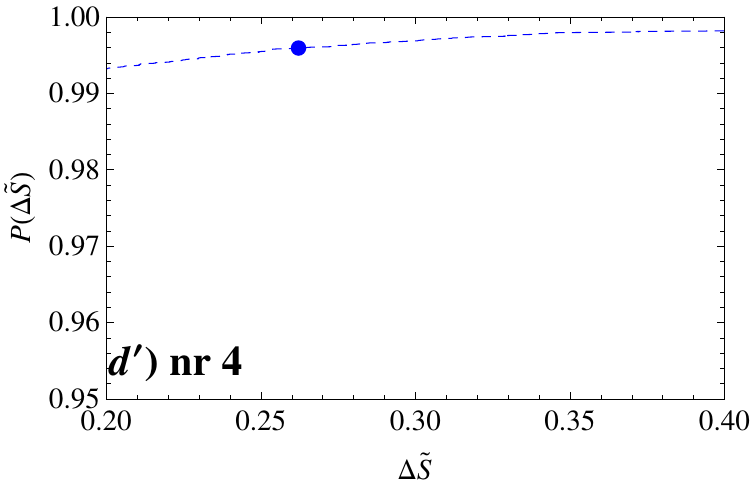}
\end{figure}
\begin{figure}[ht!]
\includegraphics[width = 0.35\linewidth]{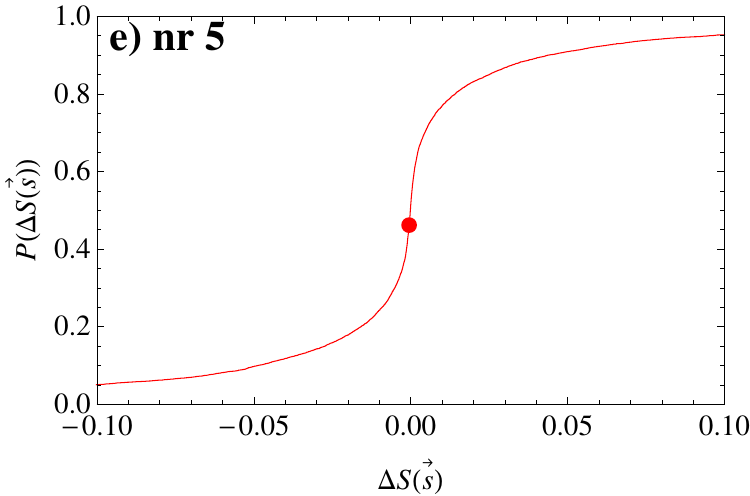}
\includegraphics[width = 0.35\linewidth]{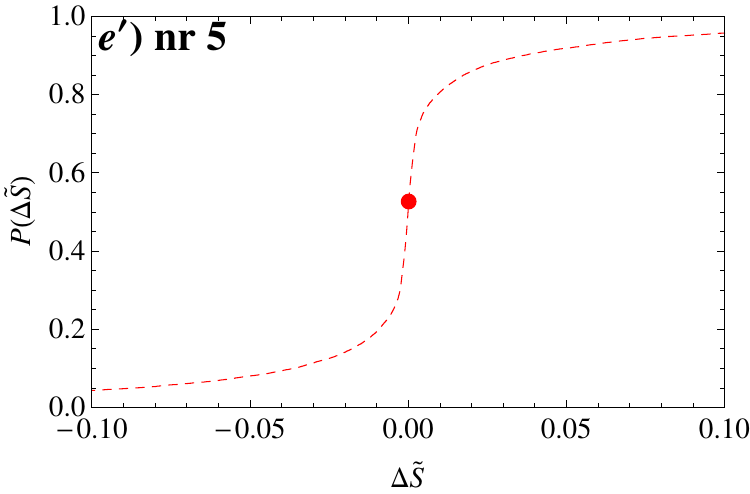}
\end{figure}
\begin{figure}[ht!]
\includegraphics[width = 0.35\linewidth]{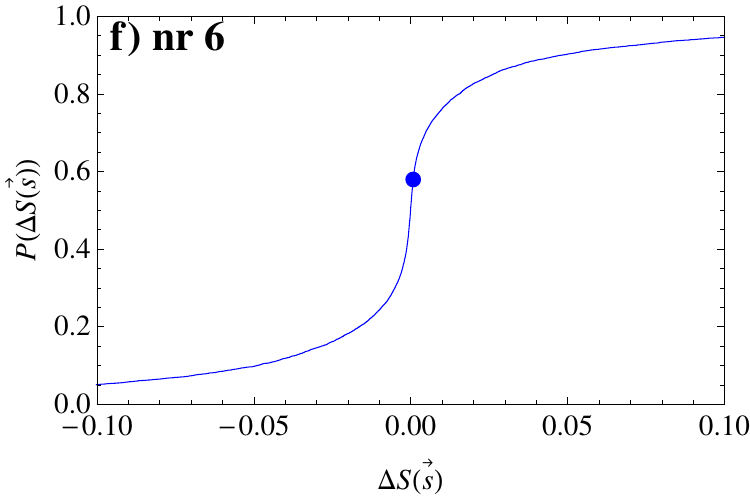}
\includegraphics[width = 0.35\linewidth]{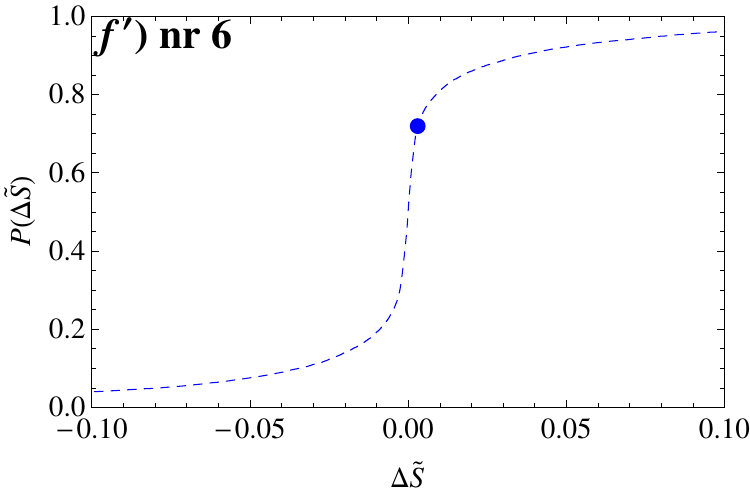}
\end{figure}
\begin{figure}[ht!]
\centering
\includegraphics[width = 0.35\linewidth]{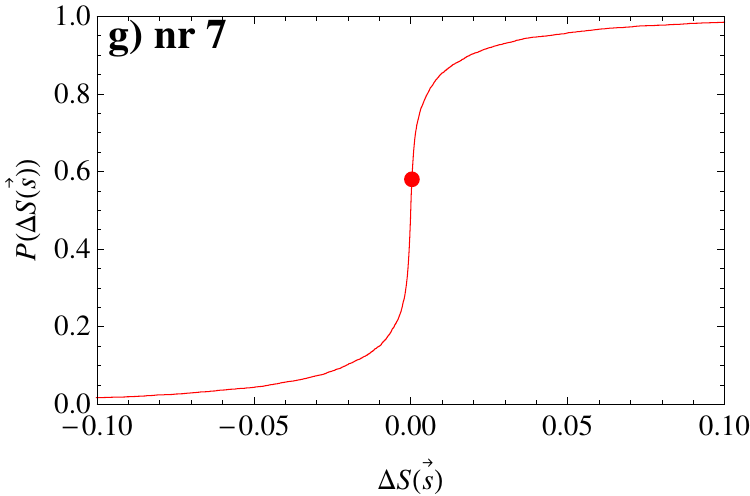}
\includegraphics[width = 0.35\linewidth]{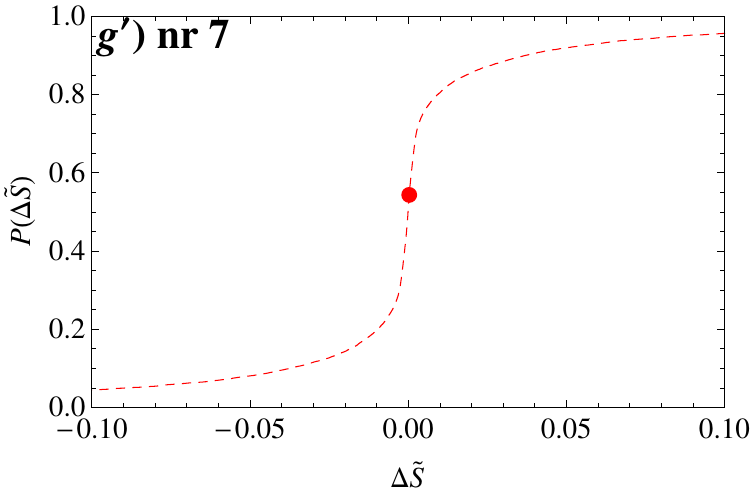}
\end{figure}
\begin{figure}[ht!]
\includegraphics[width = 0.35\linewidth]{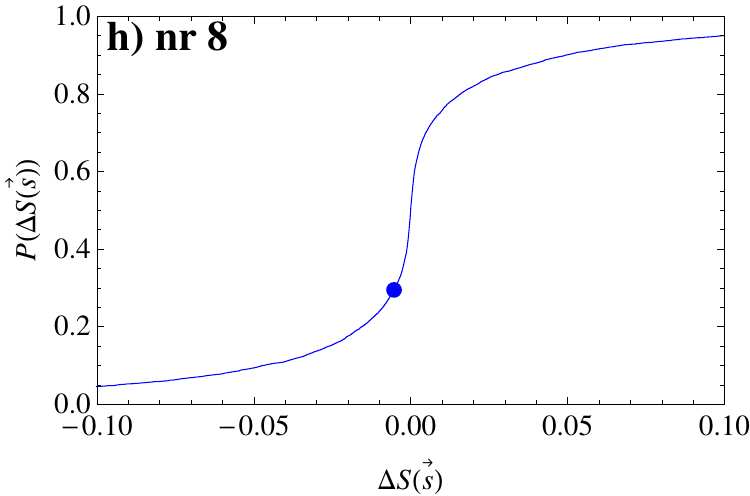}
\includegraphics[width = 0.35\linewidth]{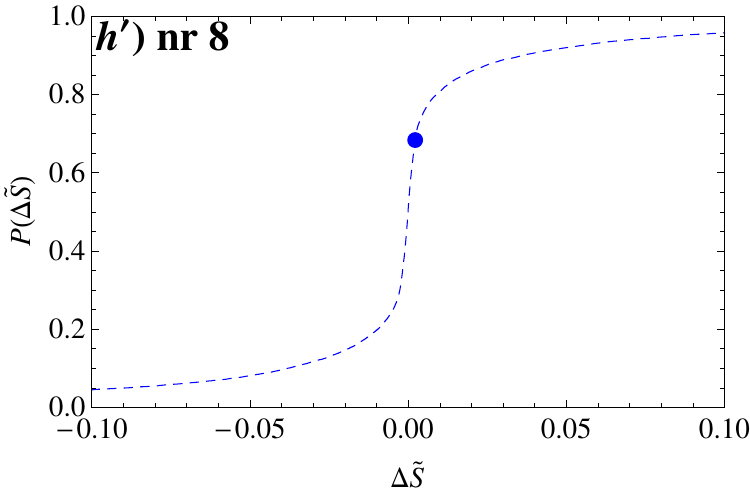}
\end{figure}
\begin{figure}[ht!]
\includegraphics[width = 0.35\linewidth]{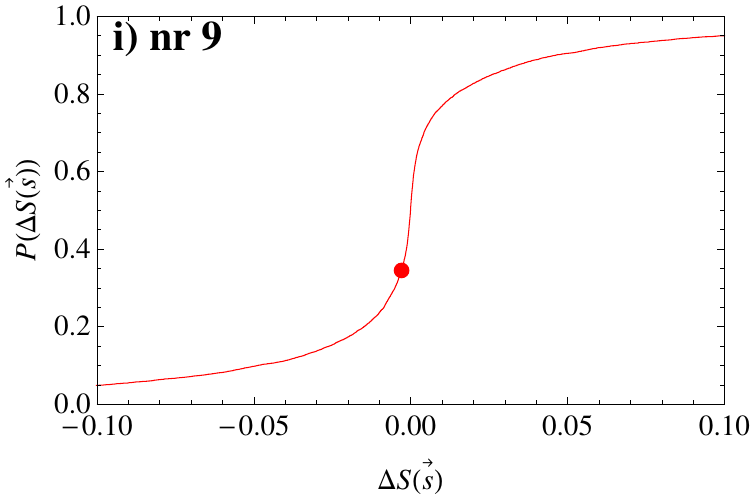}
\includegraphics[width = 0.35\linewidth]{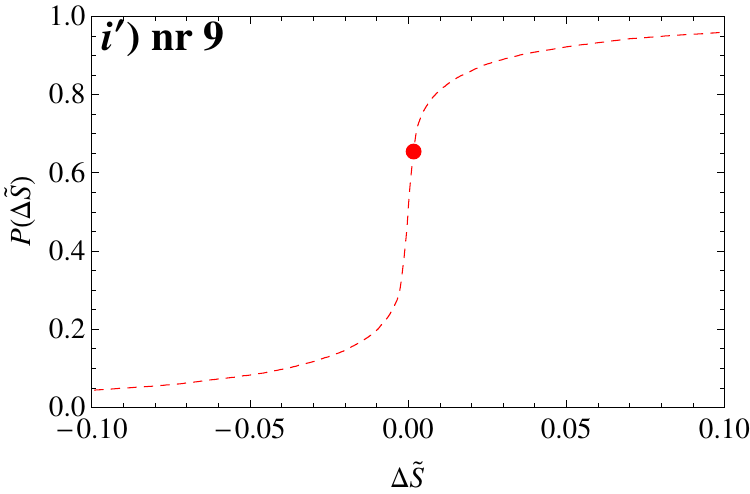}
\end{figure}
\begin{figure}[ht!]
\includegraphics[width = 0.35\linewidth]{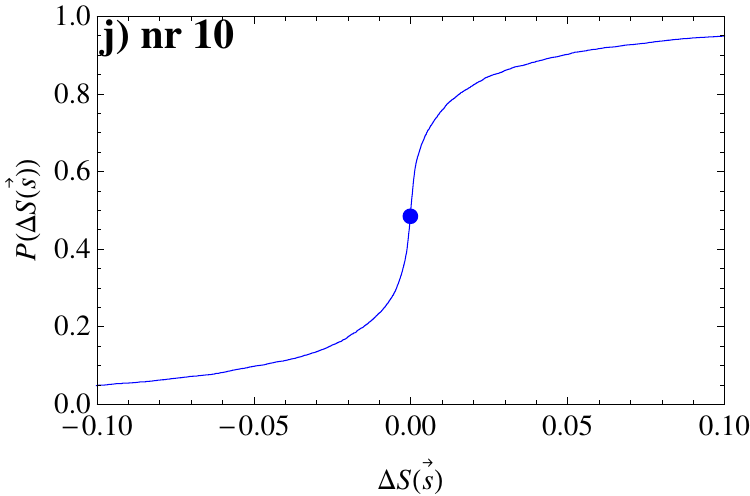}
\includegraphics[width = 0.35\linewidth]{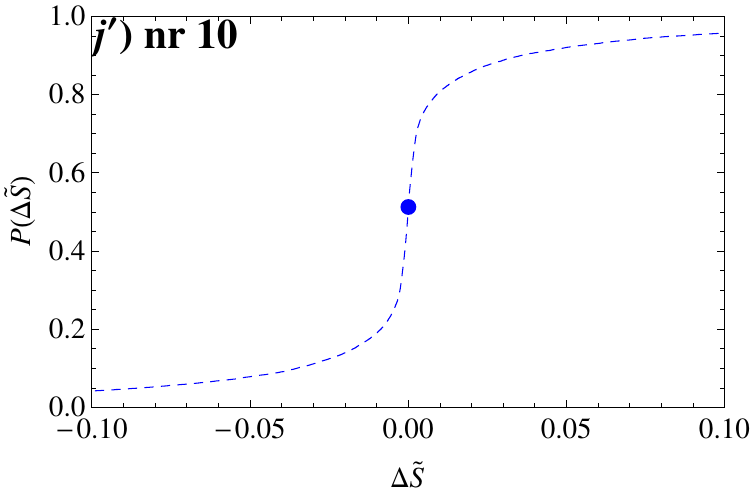}
\end{figure}
\begin{figure}[ht!]
\includegraphics[width = 0.35\linewidth]{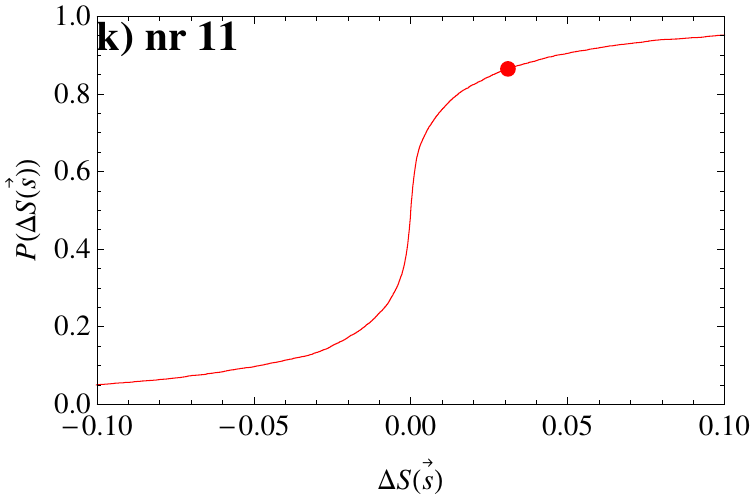}
\includegraphics[width = 0.35\linewidth]{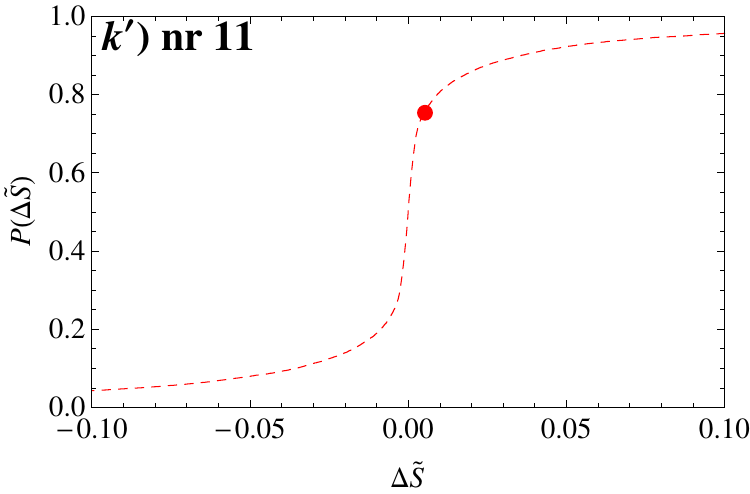}
\caption{
Cumulative distribution functions, $P(\Delta S) $, versus entropy variation. Besides the identification by the event number inn Table~I of the Main Text, we use red lines for the El Ni\~no $\rightarrow $ La Ni\~na transitions and blue lines otherwise. On the left column the plots are for the full trajectory case $\Delta S (\vec s)$ and on the right column the plots are for the case where only the extremes are considered $\Delta \tilde S (t_f, t_i)$. In each plot, the dot signals the probability of the variation of the entropy according to Table~II and Table~III presented in the Main Text.
\label{fig-cdf}}
\end{figure}

\end{widetext}

\end{document}